\documentclass[aps,prc,final,twocolumn,eqsecnum,showpacs]{revtex4-1}
\usepackage{amsmath,bm}
\usepackage{mathptmx}
 \DeclareMathAlphabet{\mathbi}{OT1}{ptm}{bx}{it}
 \DeclareSymbolFont{symbols}{OMS}{cmsy}{m}{n}
\usepackage[dvips]{graphicx}
\newif\ifonline\global\onlinefalse
\onlinetrue

\newcommand{\bp}{\mathbi{p}}
\newcommand{\br}{\mathbi{r}}
\newcommand{\BM}{\mathbf{M}}
\newcommand{\BI}{\mathrm{I}}
\newcommand{\BJ}{\mathbf{J}}

\newcommand{\cE}{\mathcal{E}}
\newcommand{\cK}{\mathcal{K}}

\newcommand{\rB}{\mathop{\rm B}}
\newcommand{\Tr}{\mathop{\rm Tr}}
\newcommand{\pp}[2]{\frac{\partial #1}{\partial #2}}

\def\<{\langle}\def\>{\rangle}

\begin{document}
\title{Periodic-orbit approach to the nuclear shell structures
with power-law potential models:\\
Bridge orbits and prolate-oblate asymmetry}
\author{Ken-ichiro Arita}
\affiliation{Department of Physics,
Nagoya Institute of Technology, Nagoya 466-8555, Japan}

%\received{29 February 2012}
%\revised{21 July 2012}
\date{9 October 2012}

\begin{abstract}
Deformed shell structures in nuclear mean-field potentials are
systematically investigated as functions of deformation and surface
diffuseness.  As the mean-field model to investigate nuclear shell
structures in a wide range of mass numbers, we propose the radial
power-law potential model, $V\propto r^\alpha$, which enables us a
simple semiclassical analysis by the use of its scaling property.  We
find that remarkable shell structures emerge at certain combinations
of deformation and diffuseness parameters, and they are closely
related to the periodic-orbit bifurcations.  In particular,
significant roles of the ``bridge orbit bifurcations'' for normal and
superdeformed shell structures are pointed out.  It is shown that the
prolate-oblate asymmetry in deformed shell structures is clearly
understood from the contribution of the bridge orbit to the
semiclassical level density.  The roles of bridge orbit bifurcations
in the emergence of superdeformed shell structures are also discussed.
\end{abstract}
\pacs{21.10.Gv, 03.65.Sq, 05.45.-a}
\maketitle

\section{Introduction}
\label{sec:intro}

Shell structures in single-particle energy spectra play essential
roles in nuclear ground-state deformations and their stabilities.
Using the semiclassical trace formula, single-particle level density
is expressed as the sum over contributions of classical periodic
orbits in the corresponding classical Hamiltonian
system\cite{Gutz,BaBlo}.  The quantum fluctuations in many-body
quantities such as energy and deformations are related to gross shell
structure in single-particle spectra determined by some short periodic
orbits.  Therefore, one can describe many-body quantum dynamics in
terms of the properties of a few important classical periodic orbits.
The single-particle shell structures are sensitively affected by
varying potential parameters such as deformations, and we have found
that bifurcations of short periodic orbits play significant roles in
emergence of remarkable shell effects.  It is a quite interesting
phenomena that the regularity of quantum spectra is enhanced by the
periodic-orbit bifurcation, which is regarded as the precursor of
chaos in classical dynamics.  In this paper, we would like to show
that the above semiclassical mechanism for the enhancement of quantum
shell effects would elucidate several problems in nuclear structure
physics.

As phenomenological mean-field potentials, modified oscillator (MO)
and Woods-Saxon (WS) models are successfully employed in shell
correction approaches.  For simpler and qualitative descriptions of
the properties of shell structures, harmonic oscillator (HO) and
infinite-well (cavity) potentials are frequently utilized for light and
heavy systems, respectively.  Axially symmetric anisotropic HO
potential models successfully explain the magic numbers of light
nuclei, emergence of superdeformed shell structures, and so on.  For
heavier nuclei, the radial profile of the potential around the nuclear
surface becomes more sharp and it looks more like a square-well
potential.  In order to avoid the complexity of treating continuum
states, the WS potential is sometimes approximated by an infinite-well
potential (cavity).  The cavity system, as well as the HO system, is
integrable under spheroidal deformation due to the existence of a
nontrivial dynamical symmetry, and several classical and quantum
mechanical quantities are obtained analytically.  It also accepts
several useful techniques to calculate quantum eigenvalue spectra,
since the Schr\"{o}dinger equation is equivalent to the Laplace
equation with Dirichlet boundary condition.

The HO and cavity systems have a significant difference in deformed
shell structures.  In the axially-deformed HO system, the ways in
which the degeneracy of levels is resolved, due to prolate and oblate
deformations, are nearly symmetric; namely, the level diagram vs
deformation is symmetric under rotation about the degenerate spherical
point by angle $\pi$.  Due to this symmetry, many-body systems between
adjacent closed-shell configurations will prefer prolate shapes in the
lower half shell and oblate shapes in the higher half shell, since
single-particle level density is lower there, and in total, prolate
and oblate shapes are expected to occur in almost the same ratios.  On
the other hand, the above kind of symmetry is apparently broken in the
cavity system.  Such asymmetry has been considered as the origin of so
called \textit{prolate-shape dominance} in nuclear ground-state
deformations: a well known experimental fact that most of the ground
states of medium-mass to heavy nuclei have prolate shapes rather than
oblate shapes.  Its origin has been discussed since the discovery of
the nuclear ground-state
deformation\cite{BM_II,Tajima96,Tajima2001,Takahara}.  This
predominance has been reproduced theoretically in microscopic
calculations.  In Hartree-Fock+BCS calculations with Skyrme
interaction\cite{Tajima96}, most of the deformed ground-state
solutions are found to have prolate shapes.  In order to pin down the
essential parameter which causes prolate-shape dominance, systematic
Nilsson-Strutinsky calculations throughout the nuclear chart have been
made\cite{Tajima2001}, and the distribution of ground-state
deformations is examined by varying the strengths of $l^2$ and $ls$
terms in the Nilsson Hamiltonian.  They found that the prolate-shape
dominance is realized under strong correlation between $l^2$ and $ls$
terms.  The recent analysis by Takahara et al. based on
Woods-Saxon-Strutinsky calculations also supports those
results\cite{Takahara}.  Hamamoto and Mottelson compared the oblate
and prolate deformation energy from the summation of single-particle
energies with spheroidal HO and cavity models, and have shown that the
prolate-shape dominance is only found in the cavity model.  They
considered the origin of the prolate-shape dominance to be the
asymmetric manner of level fannings in prolate and oblate sides which
is unique to a potential with a sharp surface, and have shown that the
above asymmetry is explained from the different roles of interaction
between single-particle levels in prolate and oblate
sides\cite{HamMot}.

We expect that the semiclassical periodic-orbit theory (POT) holds the
key for deeper understandings of above shell structures responsible
for prolate-shape dominance.  In POT, semiclassical level density is
expressed as the sum of periodic orbit (PO) contributions,
\begin{equation}
g(E)=\bar{g}(E)+\sum_\beta A_\beta(E)\cos\left(
 \frac{S_\beta(E)}{\hbar}-\frac{\pi\mu_\beta}{2}
 \right).
\label{eq:traceformula}
\end{equation}
$\bar{g}$ is the average level density equivalent to that given by the
Thomas-Fermi approximation, and the second term on the right-hand side
gives the fluctuations around $\bar{g}$.  The sum is taken over all
the classical periodic orbits $\beta$ which exist for given energy
$E$.  $S_\beta=\oint_\beta \bp\cdot d\br$ is the action integral, and
$\mu_\beta$ is the geometric phase index determined by the number of
conjugate points along the orbit.  Each orbit $\beta$ changes its size
and shape with increasing energy $E$, and the action integral
$S_\beta$ is, in general, a monotonically increasing function of $E$.
Thus, each cosine term in the PO sum (\ref{eq:traceformula}) is a
regularly oscillating function of energy whose period of oscillation
$\delta E$ is given through the relation
\begin{equation}
\delta S_\beta\sim\pp{S_\beta}{E}\delta E\sim 2\pi\hbar, \quad
\delta E\sim\frac{2\pi\hbar}{T_\beta},
\end{equation}
where $T_\beta=\partial S_\beta/\partial E$ is time period of the
orbit $\beta$.  Therefore, a gross shell structure (large $\delta E$)
is associated with short periodic orbits (small $T_\beta$).

The above fluctuation in the single-particle spectrum brings about a
fluctuation in energy of nuclei as functions of constituent nucleon
numbers.  This fluctuation part, which we call
\textit{shell energy}, is calculated by removing the smooth part from
a sum of single-particle energies by means of the Strutinsky
method\cite{Strut,BrackPauli}.  In semiclassical theory, shell energy
$E_{\rm sh}(N)$ is given by the sum of periodic orbit contribution
as\cite{BBText}
\begin{equation}
E_{\rm sh}(N)=\sum_\beta\left(\frac{\hbar}{T_\beta}\right)^2
 A_\beta\cos\left(\frac{S_\beta(E_F(N))}{\hbar}
 -\frac{\pi\mu_\beta}{2}\right), \label{eq:TF_sce}
\end{equation}
where the Fermi energy $E_F(N)$ is determined by
\begin{equation}
\int_{-\infty}^{E_F} g(E)dE=N.
\end{equation}
In Eq.~(\ref{eq:TF_sce}), the contributions of long orbits are
suppressed by the reduction factor $T_\beta^{-2}$, and the property of
shell energy is essentially determined by a few shortest periodic
orbits.  Therefore, it is sufficient to examine coarse-grained level
density where one can exclude the contribution of long periodic
orbits.

The relation between coarse-grained quantum level density oscillations
and classical periodic orbits in the spherical cavity model was first
discussed by Balian and Bloch\cite{BaBlo}.  They show that the
modulations in quantum level density oscillations are clearly
understood as the interference effect of periodic orbits with
different lengths.  This idea has been successfully applied to the
problem of supershell structure in metallic clusters\cite{Nishioka}.
Strutinsky et al.\cite{StrMag} applied periodic orbit theory
(POT)\cite{Gutz,BaBlo} to the cavity model with spheroidal deformation
and discussed the properties of deformed shell structures in
medium-mass to heavy nuclei in terms of classical periodic
orbits\cite{StrMag}.  Frisk made more extensive POT calculations to
reproduce quantum level density by the semiclassical
formula\cite{Frisk}.  He also suggested the relation between classical
periodic orbits and prolate-oblate asymmetry in deformed shell
structures, which might be responsible for the prolate-shape dominance
discussed above.  Those works have proved the virtue of semiclassical
POT for clear understanding of the properties of finite quantum
systems.

It should be emphasized here that unique deformed shell structures are
developed when the contributions of certain periodic orbits are
considerably enhanced.  The magnitude of the shell effect is related
to the amplitude factor $A_\beta$ in Eq.~(\ref{eq:traceformula}).
This amplitude factor has important dependency on the stability of the
orbit, which is generally very sensitive to the potential parameters
such as deformations.  In particular, stability factors sometimes
exhibit significant enhancement at periodic orbit bifurcations, where
new periodic orbits emerge from an existing periodic orbit.  Near the
bifurcation point, classical orbits surrounding the stable periodic
orbit form a quasiperiodic family, which makes a coherent contribution
to the level density.  This is an important mechanism for the growth
of deformed shell structures.

A typical example is the so-called superdeformed shell structure.  It
is known that single-particle spectra exhibit remarkable shell effects
at very large quadrupole-type deformation with an axis ratio around
2:1.  In the anisotropic HO model, this shell structure is related
with the periodic orbit condition; all the classical orbits become
periodic at $\omega_\perp=2\omega_z$ and they make very large
contribution to the level density fluctuation.  In the cavity model,
one also finds a significant shell effect around the 2:1 deformation,
and it is related to the bifurcations of equatorial periodic orbits
through which three-dimensional (3D) periodic orbits
emerge\cite{ASM,Magner2}.  It should be interesting to explore the
intermediate situation between the above two limits, which might
correspond to the actual nuclear situation.

Our purpose in this paper is to understand the transition of deformed
shell structure from light to heavy nuclei in terms of classical
periodic orbits.  This requires a mean field like WS potential model.
Semiclassical quantizations in spherical and deformed WS-like
potentials have been examined in Refs.~\cite{Carb,Arv}, but the
relation between classical periodic orbits and quantum level densities
has not been discussed.  As we show, the WS potential inside the
nuclear radius $R_A$ is nicely approximated by a power-law potential
which has simpler radial dependence $V\propto r^\alpha$.  This
approximation simplifies both quantum and classical calculations and
one has clear quantum-classical correspondence via the Fourier
transformation technique\cite{IJMPE}.

Thus, in the current paper, we focus on the radial dependence of the
mean-field potential (effect of surface diffuseness, described by the
$l^2$ term in the Nilsson model) and examine the shell structures
systematically as functions of deformation and surface diffuseness.
As pointed out by Tajima et al., spin-orbit coupling plays also an
important role in prolate-shape dominance.  The effect of spin-orbit
coupling will be discussed in a forthcoming paper.

This paper is organized as follows.  In Sec.~\ref{sect:model}, we
discuss the quantum and classical properties of the power-law
potential model.  The scaling properties of the model are described
and the Fourier transformation techniques are formulated.  In
Sect.~\ref{sect:sph}, quantum mechanical densities of states and shell
structures in the spherical power-law potential are examined.  Some
analytic expressions for periodic orbit bifurcations and semiclassical
formulas are given, and quantum-classical correspondence is discussed.
It will be shown that bifurcations of circular orbits bring about
unique supershell structures at several values of radial parameter
$\alpha$.  In Sec.~\ref{sect:sd}, shell structures are examined
against the spheroidal deformation parameter.  The semiclassical
origin of prolate-oblate asymmetry in deformed shell structures and
prolate-shape dominance are investigated.  The origins of
superdeformed shell structures are also examined.  Special
attention is paid to what we call ``bridge orbit bifurcations.''
Section~\ref{sect:summary} is devoted to a summary and conclusion.

\section{The power-law potential model}
\label{sect:model}

\subsection{Definition of the model}

\begin{figure}
\begin{center}
\includegraphics[width=\linewidth]{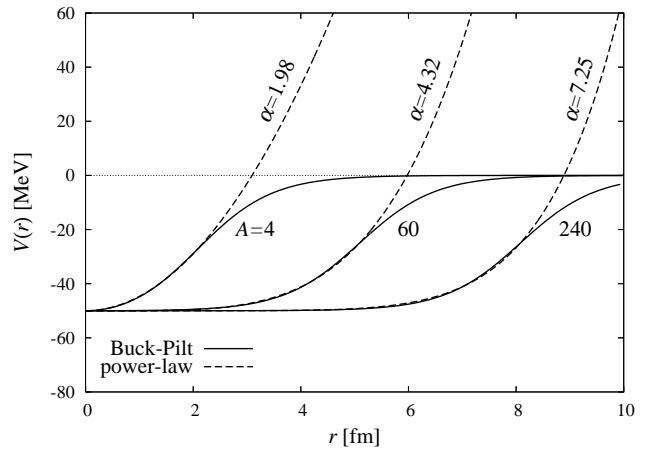}
\end{center}
\caption{\label{fig:bp}
Profiles of power-law potentials (\ref{eq:modeling})
fitted to the Buck-Pilt potentials (\ref{eq:bp}) for mass numbers
$A=4$, 60 and 240.  Values of radial parameter $\alpha$ are determined by
Eq.~(\ref{eq:msv}).}
\end{figure}

It is known that the central part of the nuclear mean-field potential
is approximately given by the Woods-Saxon (WS) model,
\begin{equation}
V_{\rm WS}(r)=-\frac{W}{1+\exp\{(r-R_A)/a\}}.
\label{eq:wsp}
\end{equation}
The depth of the potential is $W\simeq 50$~MeV, the surface
diffuseness is $a\simeq 0.7$~fm and the nuclear radius is $R_A\sim
1.3\,A^{1/3}$~fm for a nucleus with mass number $A$ \cite{Cwiok}.  The
singularity of the potential (\ref{eq:wsp}) at the origin can be
removed by replacing the WS potential with the Buck-Pilt (BP)
potential\cite{BuckPilt}
\begin{equation}
V_{\rm BP}(r)=-W\frac{1+\cosh(R_A/a)}{\cosh(r/a)+\cosh(R_A/a)}.
\label{eq:bp}
\end{equation}
By using the BP potential whose radial profile is essentially
equivalent to the WS potential, one can consider semiclassical
quantization without being concerned about the singularity in
classical orbits\cite{Carb,Arv}.  For small $A$, the inner region
($r<R_A$) of these potentials can be approximated by a harmonic
oscillator (HO).  For large $A$, these potentials are flat ($V\approx
-W$) around $r=0$ and sharply approaches zero around the surface,
looking more like a square-well potential.  In Ref.~\cite{StrMag}, the
shell energies of deformed WS potentials are compared with those for
HO and infinite square-well (cavity) potentials.  Deformed shell
structures in the WS model are similar to those of the HO model for
light nuclei, while they are more like those of the cavity model for
medium-mass to heavy nuclei.  Our aim is to understand the above
transition of deformed shell structure from the view point of
quantum-classical correspondence.  For this purpose, we take the
radial dependence of the potential as $r^\alpha$, which smoothly
connects HO ($\alpha=2$) and cavity ($\alpha=\infty$) potentials by
varying the radial parameter $\alpha$:
\begin{equation}
V_{\rm BP}(r)\approx V_\alpha(r)
=-W+\frac{W}{2}\left(\frac{r}{R_A}\right)^\alpha.
\label{eq:modeling}
\end{equation}
This \textit{power-law potential} $V_\alpha$, having a simple radial
profile, is easy to treat in both quantum and classical mechanics in
comparison with the WS/BP model.  The inner region $(r\lesssim R)$ of
the BP potential is nicely approximated by the power-law potential
(see Fig.~\ref{fig:bp}).

In Fig.~\ref{fig:bp}, the radial parameter $\alpha$ is determined so
that the power-law potential best fit the inner region $(r<R_A)$ of
the BP potential.  As a simple local matching, one may equate the
derivatives of two potentials at the nuclear surface $r=R_A$, which
gives (for $a\ll R_A$)
\begin{equation}
\alpha\sim R_A/2a.
\label{eq:alpha_lfit}
\end{equation}
Thus, the radial parameter $\alpha$ controls the surface diffuseness.
For a global fitting, we take more elaborate approach which minimizes
the volume integral of the squared potential difference inside the
nuclear radius $R_A$,
\begin{equation}
\frac{d}{d\alpha}\int_0^{R_A}dr r^2
\Bigl\{V_\alpha(r;A)-V_{\rm BP}(r;A)\Bigr\}^2 = 0.
\label{eq:msv}
\end{equation}
The value of $\alpha$ numerically obtained by Eq.~(\ref{eq:msv}) has
an approximately linear dependence on $R_A/a$,
\begin{equation}
\alpha\sim -0.62+0.68\,R_A/a, \label{eq:alpha_fit}
\end{equation}
which has qualitatively similar dependence on surface
diffuseness $a$ as the result of local fitting (\ref{eq:alpha_lfit}).

\begin{figure}
\begin{center}
\includegraphics[width=\linewidth]{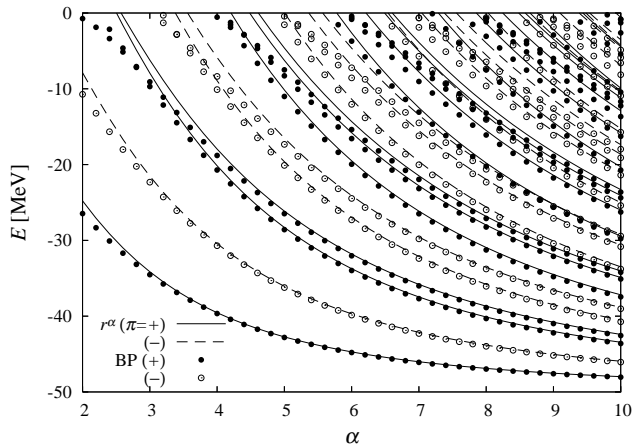}
\end{center}
\caption{\label{fig:ws_alpha}
Single-particle spectra for the spherical power-law ($r^\alpha$)
potential and the Buck-Pilt (BP) potential as functions of radial
parameter $\alpha$.  Positive and negative parity levels ($\pi=\pm)$
are respectively plotted with solid and dashed lines for the power-law
potential and with filled and open dots for the BP potential.  For the
levels of the BP model (\ref{eq:wsp}), nuclear radius $R_A$ is
determined by Eq.~(\ref{eq:alpha_fit}) as a function of $\alpha$.}
\end{figure}

Figure~\ref{fig:ws_alpha} compares single-particle level diagrams for
the BP and power-law potential models as functions of radial parameter
$\alpha$.  We use the relation (\ref{eq:alpha_fit}) to determine $R_A$
for the WS potential as a function of $\alpha$.  Although the
difference of the two potentials becomes significant at $E\gtrsim
-20~\mathrm{MeV}$, the quantum spectra for these models show fairly
nice agreements up to the Fermi energy $(E_F\sim -8~\mathrm{MeV})$ in
wide range of radial parameter $\alpha$ (see Fig.~\ref{fig:ws_alpha}).
Since most of the classical orbits have nonzero angular momentum and
they do not reach the outer bound of the potential due to centrifugal
potential, the difference of the two potentials at $r>R_A$ is hindered
in the semiclassical quantization and it might not cause much
differences in the quantum spectra up to a rather high energy.

Thus, we can employ this power-law potential model for the study of
realistic shell structures of stable nuclei from light to heavy
regions.  For unstable nuclei, the difference of the potentials at
$r>R_A$ and the effect of coupling to continuum states might become
significant.

\subsection{Scaling properties}

We have several great advantages by replacing the WS/BP potential with
the power-law potential.  The power-law potential has useful scaling
properties, which highly simplifies our semiclassical analysis.  In
the following, we eliminate the constant term $-W$ in
Eq.~(\ref{eq:modeling}) and consider the Hamiltonian for a particle of
mass $M$ moving in the deformed power-law potential as
\begin{equation}
H(\bp,\br)=\frac{p^2}{2M}
+U\left[\frac{r}{Rf(\theta,\varphi)}\right]^\alpha.
\label{eq:hamiltonian}
\end{equation}
Here, $R$ and $U$ are constants having dimension of length and energy,
respectively.  The dimensionless function $f(\theta,\varphi)$
determines the shape of the equipotential surface, and it is
normalized to satisfy the volume conservation condition
\begin{equation}
\frac{1}{4\pi}\int f^3(\theta,\varphi)d\varOmega=1,
\label{eq:volcons}
\end{equation}
which guarantees the volume surrounded by equipotential surface to be
independent of deformation.  Under a suitable scale transformation of
coordinates, the energy eigenvalue equation is transformed into a
dimensionless form,
\begin{equation}
\left[-\frac12\bm{\nabla}_u^2
 +\left(\frac{u}{f(\theta,\varphi)}\right)^\alpha\right]
 \psi(\mathbi{u})=e\psi(\mathbi{u}), \label{eq:dimlesseq}
\end{equation}
by the choice $U=\hbar^2/MR^2$ (note that the value of $U$ can be taken
arbitrarily, since the potential can be still adjusted by another parameter
$R$), and dimensionless coordinates $\mathbi{u}$ and energy $e$
defined by
\begin{equation}\label{eq:energyunit}
\mathbi{u}=\frac{\br}{R},\quad e=\frac{E}{U}\,.
\end{equation}
$\bm{\nabla}_u^2$ represents a Laplacian with respect to the
coordinate $\mathbi{u}$.  Since Eq.~(\ref{eq:dimlesseq}) does not
include constants such as $M$, $U$, $R$ and $\hbar$, one can consider
the quantum eigenvalue problem independently on those values.  Their
absolute values are determined by fitting to the BP potential through
the relation
\begin{gather*}
U\left(\frac{R_A}{R}\right)^\alpha=\frac{W}{2}.
\end{gather*}
The values of $\alpha$, $R$ and $U$ for several $A$ are listed in
Table~\ref{table:params}.

\begin{table}
\caption{\label{table:params}
Values of radial parameter $\alpha$, length unit $R$, and
energy unit $U$ of the power-law potential (\ref{eq:hamiltonian}) for
nuclei with mass number $A$.  Nuclear radius $R_A=1.3A^{1/3}$~fm,
potential depth $W=50$~MeV, surface diffuseness $a=0.7$~fm,
nucleon mass $M=938~\mathrm{MeV}/c^2$, and the
relation (\ref{eq:alpha_fit}) are used.}
\begin{center}
\begin{tabular}{c|c|c|c}\hline\hline
\hbox to5em{\hss $A$\hss} &
\hbox to5em{\hss $\alpha$\hss} &
\hbox to5em{\hss $R$~[fm]\hss} &
\hbox to5em{\hss $U$~[MeV]\hss} \\ \hline
~20 & 2.80 & 2.32 & 3.32 \\
100 & 5.23 & 3.93 & 1.14 \\
200 & 6.75 & 5.06 & 0.72 \\
\hline\hline
\end{tabular}
\end{center}
\end{table}

The scaling property of the system is particularly advantageous in the
analysis of classical dynamics.  Since the potential is a
homogeneous function of coordinates, Hamilton's equations of
motion have invariance under the following scale transformation:
\begin{equation}
(\bp,\br,t)\to
(c^{\frac12}\bp,c^{\frac1\alpha}\br,c^{\frac1\alpha-\frac12}t)
\quad \mbox{as} \quad E\to cE.
\label{eq:scaletr}
\end{equation}
Therefore, classical phase-space structure is independent of energy.
A phase-space trajectory $(\br_0(t),\bp_0(t))$ at energy $E_0$ is
transformed to a trajectory at different energy $E$ by
\begin{gather}
\br(t)=\left(\frac{E}{E_0}\right)^{\frac1\alpha}\br_0(t'), \quad
\bp(t)=\left(\frac{E}{E_0}\right)^{\frac12}\bp_0(t'), \nonumber\\
\text{with} \qquad t=\left(\frac{E}{E_0}\right)^{\frac1\alpha-\frac12}t'.
\end{gather}
Thus we have the same set of periodic orbits in an arbitrary energy
surface related through the above scale transformation.  In the
following, we set the reference energy at $E_0=U$.  The action
integral along a certain periodic orbit $\beta$ is expressed as
\begin{equation}
S_\beta(E)=\oint_{\beta(E)}\bp\cdot d\br
=S_\beta(U)\left(\frac{E}{U}\right)^{\frac12+\frac1\alpha}
\equiv\hbar\tau_\beta\cE.
\end{equation}
In the last equation, we define dimensionless ``scaled energy'' $\cE$
and ``scaled period'' $\tau_\beta$ of periodic orbit $\beta$ by
\begin{equation}
\cE=\left(\frac{E}{U}\right)^{\frac12+\frac1\alpha}, \quad
\tau_\beta=\frac{S_\beta(U)}{\hbar}.
\end{equation}
The ordinary (non-scaled) period of the orbit $\beta$ is then given by
\begin{equation}
T_\beta=\pp{S_\beta(E)}{E}
=\frac{d\cE}{dE}\hbar\tau_\beta. \label{eq:period}
\end{equation}
As one will see in the following part, it is convenient to express
periodic-orbit quantities in terms of $\cE$ and $\tau$ in place of $E$
and $T$.  In HO-type potentials ($\alpha=2$), $\cE$ and $\tau$ are
proportional to ordinary energy $E$ and period $T$, respectively.  In
cavities ($\alpha=\infty$), they are proportional to momentum $p$ and
orbit length $L$, respectively.

\subsection{Semiclassical level density}

\begin{figure}
\begin{center}
\includegraphics[width=.9\linewidth]{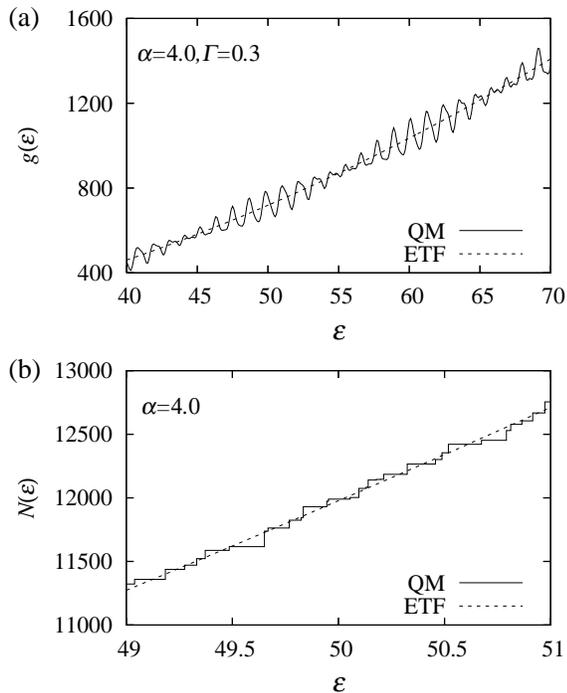}
\end{center}
\caption{\label{fig:ETF}
Comparison of quantum (QM: solid line) and semiclassical (ETF: broken
line) results for a spherical power-law potential with $\alpha=4.0$.
In panel (a), coarse-grained quantum level density (\ref{eq:gsm_qm})
with smoothing width $\varGamma=0.3$, and ETF level density
(\ref{eq:gbar_etf}) are displayed.  In panel (b), the number of
quantum levels below scaled energy $\cE$ (\ref{eq:numq}) and the ETF
average number of levels (\ref{eq:nbar_etf}) are displayed.}
\end{figure}

Let us consider the single-particle level density for the Hamiltonian
(\ref{eq:hamiltonian}).  Average level density
$\bar{g}(E)$ is given by Thomas-Fermi (TF) approximation
\begin{align}
g_{TF}(E)
&=\frac{1}{(2\pi\hbar)^3}\int d\bp\,d\br\,
   \delta(E-H(\bp,\br)) \nonumber \\
&=\frac{2\sqrt{2}}{\pi\alpha}
   \rB\left(\frac3\alpha,\frac32\right)
   \frac{\cE^3}{E}, \label{eq:gtf}
\end{align}
which is independent of deformation under volume conservation
condition (\ref{eq:volcons}).
$\rB(s,t)$ represents Euler's beta function defined by
\[
\rB(s,t)=\int_0^1 x^{s-1}(1-x)^{t-1}dx.
\]
By transforming energy $E$ to a scaled energy $\cE$, one obtains the
scaled-energy level density
\begin{equation}
g(\cE)
=\frac{dE}{d\cE}g(E)
=\frac{2\alpha}{2+\alpha}\,\frac{E}{\cE}\,g(E)
\end{equation}
Using (\ref{eq:gtf}), the average part is given by
\begin{equation}
\bar{g}(\cE)=\frac{2\sqrt{2}}{\pi}
 \rB\left(1+\frac{3}{\alpha},\frac32\right)\cE^2
\end{equation}
Correction to the TF density is obtained by the extended Thomas-Fermi (ETF)
theory\cite{Jennings,BBText},
\begin{align}
&\bar{g}_{\rm ETF}(E)=\bar{g}_{\rm TF}(E)
-\frac{1}{96\pi^2}\left(\frac{2M}{\hbar^2}\right)^{1/2}\nonumber\\
&\hspace{5em}\times\pp{}{E}
\int d\br\,\theta(E-V)\frac{\bm{\nabla}^2 V}{(E-V)^{1/2}}.
\end{align}
For the spherical case, one obtains the expression
\begin{align}
\bar{g}_{\rm ETF}(\cE)&=c_0(\alpha)\cE^2+c_1(\alpha)
\label{eq:gbar_etf} \\
&c_0(\alpha)=\frac{2\sqrt{2}}{\pi}
 \rB\left(1+\frac{3}{\alpha},\frac32\right), \nonumber\\
&c_1(\alpha)=-\frac{\alpha+1}{12\sqrt{2}\,\pi}
 \rB\left(1+\frac{1}{\alpha},\frac12\right) \nonumber
\end{align}
and the average number of levels up to scaled energy $\cE$ is given by
\begin{equation}
\bar{N}(\cE)=\frac13c_0(\alpha)\cE^3+c_1(\alpha)\cE \label{eq:nbar_etf}
\end{equation}
In Fig.~\ref{fig:ETF}, the quantum mechanically calculated coarse-grained
level density
\begin{align}
g_\varGamma(\cE)&=\int
d\cE'g(\cE')\frac{1}{\sqrt{2\pi}\varGamma}e^{-(\cE-\cE')^2/2\varGamma^2}
 \nonumber \\
 &=\sum_i\frac{1}{\sqrt{2\pi}\,\varGamma}
 e^{-(\cE-\cE_i)^2/2\varGamma^2} \label{eq:gsm_qm}
\end{align}
with smoothing width $\varGamma=0.3$ and the number of levels
\begin{equation}
N(\cE)=\sum_i\theta(\cE-\cE_i), \label{eq:numq}
\end{equation}
are compared with those in ETF approximation.  One sees that ETF (TF)
correctly describes the average properties of quantum results.  In
these plots, the differences between ETF and TF are invisibly small.

Next we consider the fluctuating part by the use of semiclassical
periodic-orbit theory\cite{BBText}.  Let us rewrite the trace formula
(\ref{eq:traceformula}) using scaled energy and scaled periods.  The
semiclassical formula for scaled-energy level density is expressed as
\begin{equation}
g(\cE)=\bar{g}(\cE)+\sum_\beta\sum_{n=1}^\infty
  A_{n\beta}(\cE)\cos\left(n\tau_\beta\cE-\frac{\pi}{2}\nu_{n\beta}
  \right). \label{eq:scaledTF}
\end{equation}
In the Hamiltonian system with no continuous symmetry, all the
periodic orbits are isolated from each other.  For a system with
continuous symmetry, e.g., a particle moving in an axially symmetric
potential, generic periodic orbits form a continuous family with respect
to the symmetry transformation, and they are called
\textit{degenerate} orbits.
For an \textit{isolated} orbit $\beta$ with $n$ repetitions, the
amplitude factor is given by the standard Gutzwiller
formula\cite{Gutz,BBText},
\begin{equation}
A_{n\beta}=\frac{T_\beta}{\pi\hbar\sqrt{|\det(\BI-\BM_\beta^n)|}}
 \,\frac{dE}{d\cE}
 =\frac{\tau_\beta}{\pi\sqrt{|\det(\BI-\BM_\beta^n)|}}. \label{eq:A_Gutz}
\end{equation}
In the last equation, we used Eq.~(\ref{eq:period}).  $\BM_\beta$
represents the monodromy matrix\cite{Baranger,BBText},
which is a linearized Poincar\'e map defined by
\begin{equation}
\BM_\beta=\pp{(\br_\perp(T_\beta),\bp_\perp(T_\beta))}{(\br_\perp(0),
 \bp_\perp(0))}\,,
\end{equation}
where $(\br_\perp(t),\bp_\perp(t))$ are the local coordinates and momenta
perpendicular to the periodic orbit $\beta$ as functions of time $t$,
and $T_\beta$ is the period of the primitive orbit.

In a two-dimensional
autonomous Hamiltonian system, monodromy matrix $\BM$ is a $(2\times
2)$ real and symplectic matrix,
\[
\BM\,\BJ\,\BM^T=\BJ, \quad \BJ=\begin{pmatrix}0 & 1\\ -1 & 0\end{pmatrix},
\]
and its eigenvalues appear in one of the following three
forms\cite{BBText,Baranger}:
\begin{enumerate}\def\labelenumi{(\alph{enumi})}
\item $(e^u,e^{-u})$: hyperbolic with no reflection,
$\Tr\BM=2\cosh u > 2$,
\item $(e^{iv},e^{-iv})$: elliptic,
$\Tr\BM=2\cos v$, $|\Tr\BM|\leq 2$,
\item $(-e^u,-e^{-u})$: hyperbolic with reflection,
$\Tr\BM=-2\cosh u<-2$.
\end{enumerate}
The orbit is stable in case (b) and otherwise unstable, and the
stability of the orbit is determined by the trace of monodromy matrix.
The stability factor in Eq.~(\ref{eq:A_Gutz}) is also determined by the
trace of the monodromy matrix:
\begin{equation}
\det(\BI-\BM_\beta)=2-\Tr\BM_\beta.
\end{equation}
The eigenvalues of $\BM$ (and therefore $\Tr\BM$) are independent of a
choice of Poincar\'e surface or a choice of canonical variables.
These eigenvalues continuously vary as deformation changes, and it
happens that they become unity at certain values of deformation,
namely, $u=0$ in (a) or $v=0$ in (b).  At those deformations,
the Poincar\'e map acquires a new fixed point in the direction of
eigenvector $\delta Z_1$ belonging to the unit eigenvector:
\begin{equation}
\BM\delta Z_1=\delta Z_1.
\end{equation}
In this way, periodic orbit bifurcation occurs at $\Tr\BM=2$.  The
number of new emerging orbits is dependent on the type of the
bifurcation\cite{Ozorio}.  When a stable (unstable) orbit undergoes
{\it pitchfork} bifurcation, it turns unstable (stable) and a new
stable (unstable) orbit emerges from it.  When a stable orbit
undergoes {\it period-doubling} bifurcation, a pair of stable and
unstable orbits will emerge.

In a three-dimensional Hamiltonian system, the size of the monodromy
matrix becomes $(4\times 4)$.  Under axial symmetry, periodic orbits
degenerate with respect to the rotation, and the monodromy matrix has
unit eigenvalue corresponding to the direction of the rotation.  Thus,
by removing the rotational degrees of freedom, the stability of the
orbit is described by a $(2\times 2)$ symmetry-reduced monodromy matrix,
and it has the same properties as in the two-dimensional case.
For such degenerate orbits in the system with continuous
symmetry, the trace formula is modified by what is called
\textit{extended Gutzwiller theory}\cite{Creagh,BBText}.  The
amplitude factor for the degenerate orbit is proportional to the
stability factor similar to that in (\ref{eq:A_Gutz}), but with
symmetry-reduced monodromy matrix $\tilde{\BM}_\beta$.
For fully degenerate orbits in an integrable system, one can use
the Berry-Tabor formula\cite{BerTab}.

In general, the stability factor
$|\det(\BI-\tilde{\BM}_\beta^n)|^{-1/2}$ has strong dependence on the
deformation parameter, and is responsible for the sensitivity of shell
structures to deformations.  The divergence of the Gutzwiller
amplitude (\ref{eq:A_Gutz}) based on the standard stationary phase
method can be remedied by improved treatment of the trace integral in
phase space (e.g., uniform approximations\cite{Sieber96,SS97,SS98} and
the improved stationary-phase method\cite{Magner1,Magner2}) and one can
obtain finite amplitudes through the bifurcation processes.  Those
amplitudes sometimes show strong enhancement around the bifurcation
points, since the monodromy matrix has a unit eigenvalue there, and a
local family of quasi-periodic orbits is formed in the direction of the
eigenvector $\delta Z_1$ belonging to the unit eigenvalue, which make
a coherent contribution to the level density.

One should, however, note that the above enhancement is not always
found for every bifurcations.  The significance of bifurcation depends
on the normal form parameters which describe nonlinear dynamics around
the periodic orbit at the bifurcation points.  In Ref.~\cite{Kaidel},
uniform approximation remedies the divergence problems which one
encounters at bifurcation points in the standard stationary phase
method, but the obtained amplitude show no enhancement around there.
In Ref.~\cite{ABtri}, we found very strong enhancement of amplitude
around the bifurcation point for one certain orbit, but the same type
of bifurcation in another orbit shows no enhancement.  In our previous
studies, we have shown that significant growth of shell effects at a
certain deformation is related with bifurcations of
\textit{simple short} periodic orbits\cite{SAM,ASM,Magner1,Magner2,ABtri}.

\subsection{Fourier transformation technique}\label{sec:fourier}

The Fourier transformation technique is especially useful in studying
classical-quantum correspondence in the system with scale invariance.
Let us consider the Fourier transform of scaled-energy level density
\begin{equation}
F(\tau)=\int g(\cE)e^{i\tau\cE}e^{-\frac12(\gamma\cE)^2}d\cE.
\label{eq:fourier}
\end{equation}
In the integrand, Gaussian damping factor is included in order to
exclude the level density at high energy $\gamma\cE\gg 1$ where
the numerically obtained single-particle spectra do not have good
precision.

By inserting the quantum level density
$g(\cE)=\sum_n\delta(\cE-\cE_n)$ into Eq.~(\ref{eq:fourier}), one
obtains
\begin{equation}
F^{\rm qm}(\tau)=\sum_{\cE_n<\cE_{\rm max}}
 e^{i\tau\cE_n-\frac12(\gamma\cE_n)^2}, \label{eq:fourier_qm}
\end{equation}
which can be easily evaluated using quantum mechanically calculated
energy eigenvalues $\{\cE_n\}$.  On the other hand, by inserting the
semiclassical level density (\ref{eq:scaledTF}), one formally has the
expression
\begin{equation}
F^{\rm cl}(\tau)=\bar{F}(\tau)
+\pi\sum_{n\beta}e^{i\pi\mu_{n\beta}/2}A_{n\beta}(-i\partial_\tau)
\delta_\gamma(\tau-n\tau_\beta). \label{eq:ftl_cl}
\end{equation}
Here, $\delta_\gamma(z)$ represents a normalized Gaussian with
width $\gamma$
\begin{equation}
\delta_\gamma(z)=\frac{1}{\sqrt{2\pi}\,\gamma}\,
e^{-\frac{z^2}{2\gamma^2}},
\end{equation}
which coincides with Dirac's delta function in the limit $\gamma\to
0$.  Thus, $F(\tau)$ should be a function possessing successive peaks
at the scaled periods of classical periodic orbits $\tau=n\tau_\beta$.
[In Eq.~(\ref{eq:ftl_cl}), the argument $\cE$ of the amplitude
$A(\cE)$ is formally replaced with differential operator
$-i\partial_\tau$.  For an isolated orbit, the amplitude is a constant
and the corresponding term in Eq.~(\ref{eq:ftl_cl}) becomes a simple
Gaussian.  For a degenerate family of degeneracy $\cK$, the amplitude
is proportional to $\cE^{\cK/2}$ and the peak might not be exactly
centered at the scaled action.]  Therefore, by calculating the Fourier
transform of the quantum mechanical level density, one can extract
information on the significance of each periodic orbits contributing
to the semiclassical level density.  The parameter $\gamma$ implies the
resolution of the periodic orbit in the Fourier transform.  For a
better resolution, a larger number of good quantum energy levels (up to
$\cE_{\rm max} \gtrsim 2/\gamma$) are required.

\section{Spherical power-law potentials}
\label{sect:sph}

\subsection{Classical periodic orbits}

In the spherical power-law potential model, several simple analytic
descriptions for the properties of the periodic orbits are available.
In the following, we take the units $\hbar=M=R=U=1$ for simplicity.
Taking the orbits in the $(x,y)$ plane and setting the $z$ component of
the angular momentum to $l_z=K$, the two-dimensional effective
Hamiltonian in polar coordinates is written as
\begin{equation}
H=\frac12 p_r^2+V_{\rm eff}(r;K), \quad
V_{\rm eff}(r;K)=r^\alpha+\frac{K^2}{2r^2}
\label{eq:H_polar}
\end{equation}
The circular orbit $r(t)=r_c$ (denoted by C) satisfies the condition
\begin{equation}
\left(\pp{V_{\rm eff}}{r}\right)_{r_c}=0,
\end{equation}
from which one obtains, for energy $E$,
\begin{equation}
r_c=\left(\frac{2E}{2+\alpha}\right)^{1/\alpha},
\end{equation}
and the angular frequency
\begin{equation}
\omega_c=\frac{K}{r_c^2}
=\sqrt{\alpha}\left(
\frac{2E}{2+\alpha}\right)^{\frac12-\frac1\alpha}.
\end{equation}
Thus, the scaled period of the orbit C is analytically given by
\begin{equation}
\tau_{\rm C}=2\pi\sqrt{\alpha}\left(\frac{2}{2+\alpha}\right)^{\frac{1}{2}
 +\frac{1}{\alpha}}.
\end{equation}
The circular orbit is stable, and $r(t)$ of the orbits in vicinities of
the circular orbit oscillate
around $r_c$ with angular frequency
\begin{equation}
\varOmega_c
=\sqrt{\left(\pp{^2V_{\rm eff}}{r^2}\right)_{r_c}}
=\sqrt{\alpha(\alpha+2)}\left(
\frac{2E}{2+\alpha}\right)^{\frac12-\frac1\alpha}.
\end{equation}
Bifurcations occur when the ratio of those two frequencies $\omega_c$
and $\varOmega_c$ becomes rational, namely,
\begin{equation}
\frac{\Omega_c}{\omega_c}=\sqrt{\alpha+2}=\frac{n}{m}
\end{equation}
for \textit{period $m$-upling} bifurcation.  Here, a new orbit which
oscillates $n$ times in the radial direction when it rotates $m$ times
along the orbit C emerges from $m$th repetition of orbit C.  The
values of $\alpha$ at such bifurcations are given by
\begin{align}
\alpha &=\frac{n^2}{m^2}-2 \label{eq:bifpts} \\
&=\left\{
\begin{array}{l@{\quad}l}
 2,7,14,\cdots &
 ,m=1 \\[2pt]
 \left(\dfrac{8}{4}\right),\dfrac{17}{4},
 \left(\dfrac{28}{4}\right),\dfrac{41}{4},\cdots &
 ,m=2 \\[8pt]
 \left(\dfrac{18}{9}\right),\dfrac{31}{9},\dfrac{46}{9},
 \left(\dfrac{63}{9}\right),\dfrac{82}{9},\dfrac{103}{9},\cdots &
 ,m=3 \\[5pt]
 \multicolumn{1}{c}{\vdots} & \multicolumn{1}{c}{\vdots}
\end{array}\right. \nonumber
\end{align}
The numbers in parenthesis are those which already appeared in smaller
$m$, corresponding to the repetitions of primitive orbits.

\begin{figure}
\begin{center}
\includegraphics[width=.8\linewidth]{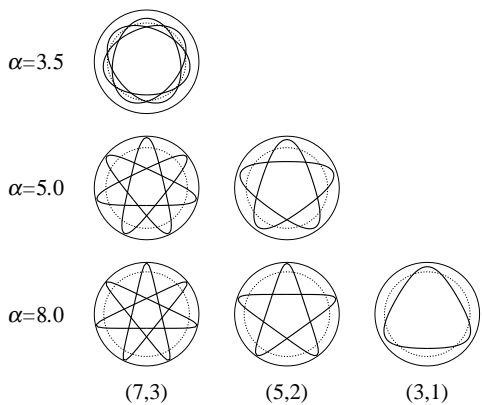}
\end{center}
\caption{\label{fig:orbit_sph}
Some short periodic orbits $(n,m)$ in spherical power-law potentials.
They emerge from the circular orbit (shown with a broken line) via
period $m$-upling bifurcations.  The outermost circle represents the
boundary of classically accessible region.}
\end{figure}

Figure~\ref{fig:orbit_sph} shows some periodic orbits $(n,m)$ emerging
from the circular orbit via period $m$-upling bifurcations of circular
orbit C.
In the following subsections, we will show that
the above bifurcations bring about unique shell structures due to the
interference of shortest orbit and bifurcated orbits which
manifest at certain values of radial parameter $\alpha$.

Another periodic orbit is the diameter orbit denoted by X.
The scaled period of the orbit X is also given analytically;
\begin{equation}
\tau_{\rm X}=2\sqrt{2}\rB\left(1+\frac{1}{\alpha},\frac12\right).
\end{equation}
In the limit $\alpha\to 2$, the orbits X and C make degenerate family
with scaled period $\tau=\sqrt{2}\pi$.  The diameter orbits in
spherical potential cause no bifurcations by varying $\alpha$.

\subsection{Fourier analysis of quantum level density}

\begin{figure}[b]
\begin{center}
\ifonline
\includegraphics[width=\linewidth]{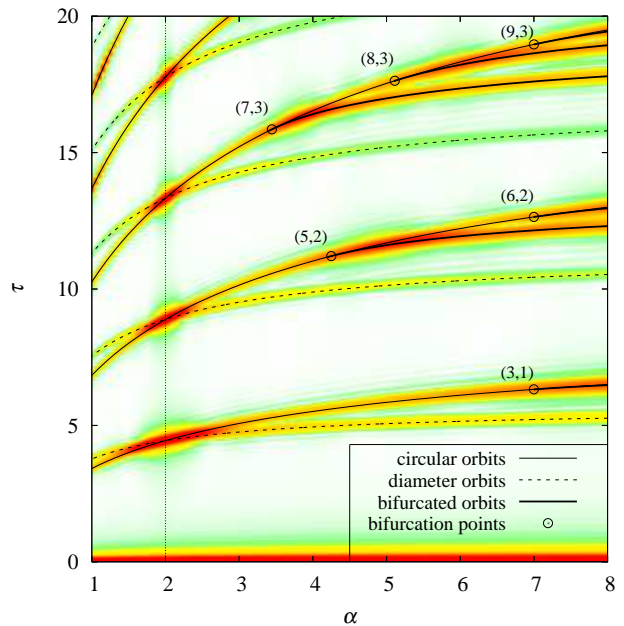}
\else
\includegraphics[width=\linewidth]{Fig05.ps}
\fi
\end{center}
\caption{\label{fig:ftlmap_sph}
(Color online) Gray-scale plot of the Fourier transform of quantum
level density (\ref{eq:fourier_qm}) as a function of radial parameter
$\alpha$ and scaled period $\tau$.  The modulus of the Fourier
transform $|F(\tau)|$ has a large value in the dark region.  Scaled
periods of classical periodic orbits $\tau_\beta(\alpha)$ are also
shown with lines as functions of $\alpha$.  Bifurcation points $(n,m)$
given by Eq.~(\ref{eq:bifpts}) are indicated by open circles.}
\end{figure}

As discussed in Sec.~\ref{sec:fourier}, the Fourier transform of 
scaled-energy level density $g(\cE)$ will exhibit peaks at the
scaled periods $n\tau_\beta$ of classical periodic orbits.
We calculate the quantum spectra by smoothly varying the radial
parameter $\alpha$, and take the Fourier transform of the level density
for each $\alpha$.  The Fourier amplitude as a function of $\alpha$ and
scaled period $\tau$ is shown in Fig.~\ref{fig:ftlmap_sph}.
The scaled periods of classical periodic orbits are also drawn as
functions of $\alpha$.  One finds an excellent correspondence between
Fourier peaks and classical periodic orbits.  The peak at $\tau=0$
corresponds to the average level density, which in semiclassical
theory is derived from the contribution of the
\textit{zero-length} or \textit{direct} trajectory. Equally spaced
remarkably large peaks for $\alpha=2$ are of a fully degenerate periodic
orbit family (and its repetitions) in an isotropic harmonic oscillator
[limit of SU(3) symmetry].  If the $\alpha$ is slightly shifted from
this value, the orbit family bifurcates into circular orbit and
diametric orbit families.  With increasing $\alpha$, the circular orbit
and its repetitions encounter successive bifurcations at the values
given by Eq.~(\ref{eq:bifpts}).  One will see that the Fourier peaks
associated with the orbits are strongly enhanced around those
bifurcation points, indicated by open circles in
Fig.~\ref{fig:ftlmap_sph}.  This clearly illustrates the significance of
periodic-orbit bifurcations to the enhancement of shell effect.  One
will also note that the maxima of the Fourier amplitudes are slightly
shifted towards post-bifurcation side as a general trend.  Such
shifts have been explained in the semiclassical theory, which is
extended to be able to treat bifurcations, e.g., in the improved
stationary-phase method\cite{Magner1,Magner2} and in the uniform
approximation\cite{ABbridge}.

\begin{figure}
\begin{center}
\includegraphics[width=\linewidth]{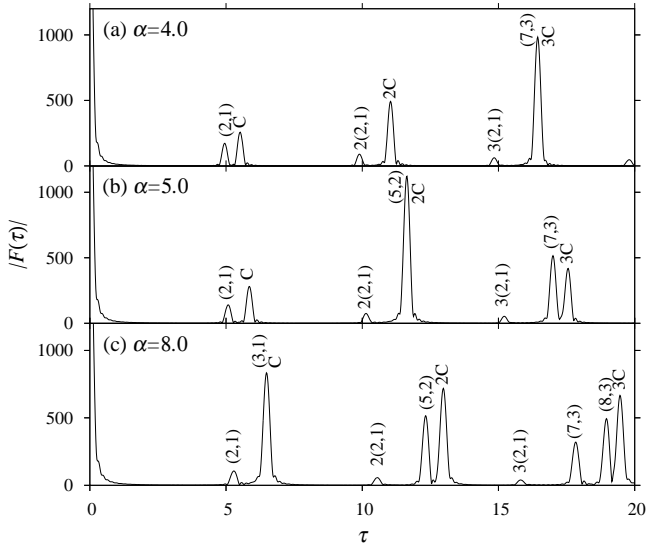}
\end{center}
\caption{\label{fig:ftl_sph}
Moduli of Fourier transforms of quantum level density
(\ref{eq:fourier_qm}) plotted as functions of scaled period $\tau$ for
(a) $\alpha=4.0$, (b) 5.0, and (c) 8.0.  The peaks associated with the
periodic orbits are labeled by their indices $(n,m)$.  (2,1) represents
a diametric orbit, C represent a circular orbit, and $k(n,m)$ represents
the $k$th repetition of the primitive orbit $(n,m)$.  In panel (a),
the scaled periods of (7,3) and 3C orbits are so closed that the
Fourier transform is not resolved into their individual peaks; the same
for (5,2) and 2C orbits in panel (b), and for (3,1) and C orbits in
panel (c).}
\end{figure}

\subsection{Bifurcation enhancement effect to the shell structures}

\begin{figure}
\begin{center}
\ifonline
\includegraphics[width=\linewidth]{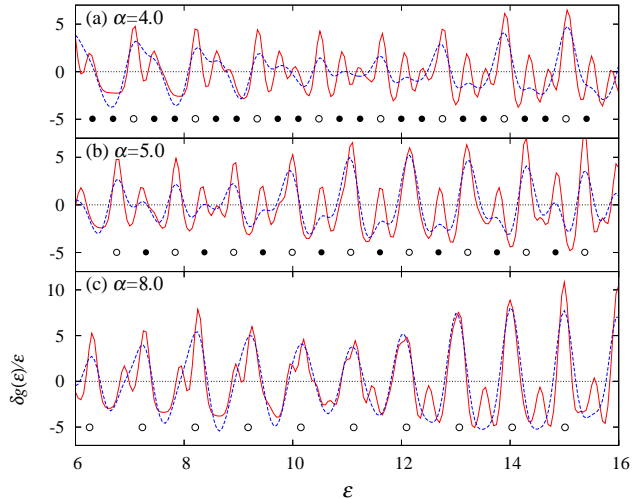}
\else
\includegraphics[width=\linewidth]{Fig07.ps}
\fi
\end{center}
\caption{\label{fig:sld_sph}
(Color online) Oscillating part of the coarse-grained scaled-energy
level density $\delta g(\cE)$ divided by $\cE$ in the spherical power-law
potential model with (a) $\alpha=4.0$, (b) 5.0 and (c) 8.0.  Solid and
dashed lines show results with smoothing width $\varGamma=0.12$ and
0.24, respectively.  In panel (a), dots are placed with interval
$\delta\cE=0.38$, which approximately coincide with the positions of
level density maxima.  The level density takes relatively larger
values at the open dots.  The dots in panels (b) and (c) are placed
with intervals $\delta\cE=0.54$ and 0.97, respectively, which also
coincide with the positions of level density maxima. (Physically, the
level density minima have more significance, but the supershell
structures are clearer for the maxima in these plots.)}
\end{figure}

In order to see the effect of bifurcations of orbits (7,3), (5,2), and
(3,1), we examine the shell structures at $\alpha=4.0$, 5.0, and 8.0
where the Fourier amplitudes corresponding to the above orbits are
most enhanced.  Figure~\ref{fig:ftl_sph} shows the moduli of Fourier
amplitudes $|F(\tau)|$ for the above values of radial parameter $\alpha$
(the cross sectional view in Fig.~\ref{fig:ftlmap_sph} along the vertical
lines at those values of $\alpha$).
Figure~\ref{fig:sld_sph} shows the oscillating part of the coarse-grained
level densities with two choices of smoothing parameter,
$\varGamma=0.24$ for extracting only the gross shell structures and
$\varGamma=0.12$ for additional finer structures.

In Fig.~\ref{fig:ftl_sph}(a), one sees the largest (except for
$\tau=0$) peak at $\tau\sim 16.5$, which corresponds to the third
repetitions of the circular orbit, 3C ($3\tau_{\rm C}=16.539$), as
well as orbit (7,3) ($\tau_{7,3}=16.410$) bifurcated from 3C at
$\alpha=31/9=3.44$.  These orbits are expected to make dominant
contributions in the periodic orbit sum (\ref{eq:scaledTF}), and the
pitch of the level density oscillation should be given by
$\delta\cE=2\pi/\tau_\beta\approx 0.38$.  The oscillating level
density shown in Fig.~\ref{fig:sld_sph}(a) has the period of
oscillation just as predicted above.  One also note that the
oscillation is regularly modulated and the amplitude becomes
relatively large for every three oscillations.  This is a typical
supershell structure caused by the interference of period-3 and
period-1 orbits.

\begin{figure}
\begin{center}
\includegraphics[width=\linewidth]{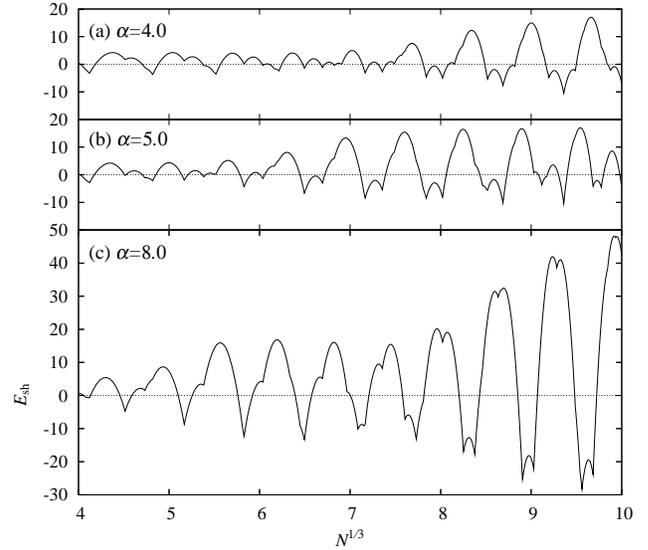}
\end{center}
\caption{\label{fig:sce_sph}
Shell energy for (a) $\alpha=4.0$, (b) 5.0, and (c) 8.0,
as functions of the cubic root of particle number, taking
account of the spin degeneracy factor.}
\end{figure}

In Fig.~\ref{fig:ftl_sph}(b), one sees a prominent peak at $\tau\simeq
11.6$ associated with the second repetitions of the circular orbit, 2C
($2\tau_{\rm C}=11.690$), as well as orbit (5,2) ($\tau_{5,2}=11.609$)
bifurcated from 2C at $\alpha=4$.  The contribution of these orbit to
the level density should be the oscillating function of scaled energy
$\cE$ with the period $\delta\cE\approx 0.54$.  The oscillating level
density shown in Fig.~\ref{fig:sld_sph}(b) has just the same period as
predicted above.  One also notes that the supershell structure caused by the
interference of period-2 and period-1 orbits is manifested.

In Fig.~\ref{fig:ftl_sph}(c), one sees a large peak at $\tau\sim 6.5$
associated with the primitive circular orbit C ($\tau_{\rm C}=6.499$)
and the orbit (3,1) ($\tau_{3,1}=6.479$) bifurcated from C at
$\alpha=7$.  The contributions of these orbits bring about the
oscillation of the level density with period $\delta\cE\approx 0.97$.  The
calculated quantum level density in Fig.~\ref{fig:sld_sph}(c) shows
the behavior just as expected.

The above shell and supershell structures are also reflected in the
shell energy shown in Fig.~\ref{fig:sce_sph}.  In panels (a) and
(b), the subshell structures due to period-3 and period-2 orbits,
respectively, can be found for large $N$ ($N^{1/3}\gtrsim 6$),
although they are not so evident in comparison with those found in the
level density due to the reduction factor $T_\beta^{-2}$ in
the trace formula of shell energy (\ref{eq:TF_sce}).  In panel (c)
of Fig.~\ref{fig:sce_sph}, one sees a remarkable enhancement of major
shell effects compared with the other panels.  This is regarded as the
result of bifurcation enhancement effect of the circular orbit.
Note that the plots in Figs.~\ref{fig:sld_sph} and \ref{fig:sce_sph}
are extended to large $\cE$ and $N$ (far beyond the region of existing
nuclei, but this may be meaningful for metallic clusters), where the
above shell and supershell structures become more evident.
Unfortunately, the subshell structures for $\alpha=4.0$ and 5.0 in
shell energies are not very prominent in the existing nuclear region
and they might disappear, e.g., after including the spin-orbit
coupling, but the pronounced shell effect for $\alpha\sim 8.0$ might
survive and be responsible for enhancement of shell effects in real
nuclei around the medium-mass to heavy region.

\section{Spheroidal deformations}
\label{sect:sd}

\subsection{Shape parametrization and quantum spectra}

\begin{figure}
\includegraphics[width=\linewidth]{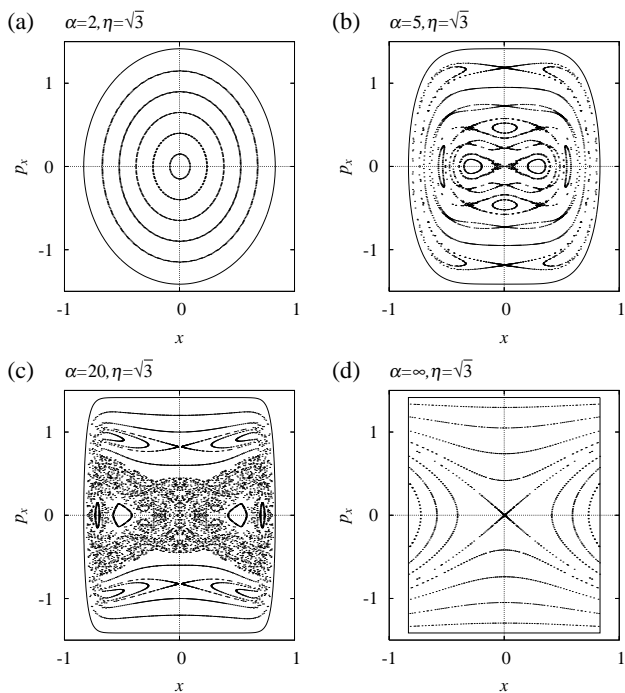}
\caption{\label{fig:poincare}
Poincar\'e surface of section $\{(x,p_x)|z=0\}$ for meridian-plane orbits
in spheroidal-shape potentials with axis ratio
$\eta=\sqrt{3}$ ($\delta\approx 0.49$) and with several values of
radial parameter $\alpha$.
The origin $(x=p_x=0)$ corresponds to the orbit Z, and the outer
boundary corresponds to the orbit X.}
\end{figure}

An axially symmetric anisotropic harmonic oscillator potential system
is integrable, and it has a spheroidal equipotential surface.
It is known that a spheroidal deformed cavity (infinite well potential)
system is also integrable.  For spheroidal deformation, the shape function
is expressed as
\begin{equation}
f(\theta)
% =\sqrt{\frac{r^2/R_0^2}{\dfrac{x^2+y^2}{R_\perp^2}
% +\dfrac{z^2}{R_z^2}}} \nonumber\\
=\left[\frac{\sin^2\theta}{(R_\perp/R_0)^2}
 +\frac{\cos^2\theta}{(R_z/R_0)^2}\right]^{-1/2}
\label{eq:deform_sph}
\end{equation}
where $R_z$ and $R_\perp$ represent lengths of semiaxes of the
spheroid which are parallel and perpendicular to the symmetry axis
($z$ axis), respectively, and $R_0$ is their spherical value.
Taking account of the volume conservation condition
$R_\perp^2R_z=R_0^3$, we define the deformation parameter
$\delta$ as
\begin{equation}
R_\perp=R_0e^{-\delta/3}, \quad R_z=R_0e^{2\delta/3}.
\label{eq:delta}
\end{equation}
It is related to the axis ratio $\eta=R_z/R_\perp$ by $\eta=e^\delta$.
The spherical shape $\eta=1$ corresponds to $\delta=0$ and
prolate/oblate superdeformed shapes $\eta=2^{\pm 1}$ correspond to
$\delta=\pm\ln 2\approx\pm 0.69$.  The system with spheroidal
power-law potential is nonintegrable except for two limits, $\alpha=2$
(HO) and $\alpha=\infty$ (cavity).  In Fig.~\ref{fig:poincare}, we
show the Poincar\'e surface of section for $\alpha=2,5,20$, and $\infty$,
each with $\eta=\sqrt{3}$ ($\delta\approx 0.49$).  It is found that
some complex structures emerge in the Poincar\'e plots with increasing
$\alpha>2$, and the surface becomes most chaotic around $\alpha\sim
20$, then it turns into simpler structure for extremely large
$\alpha$.

\begin{figure}
\begin{center}
\includegraphics[width=\linewidth]{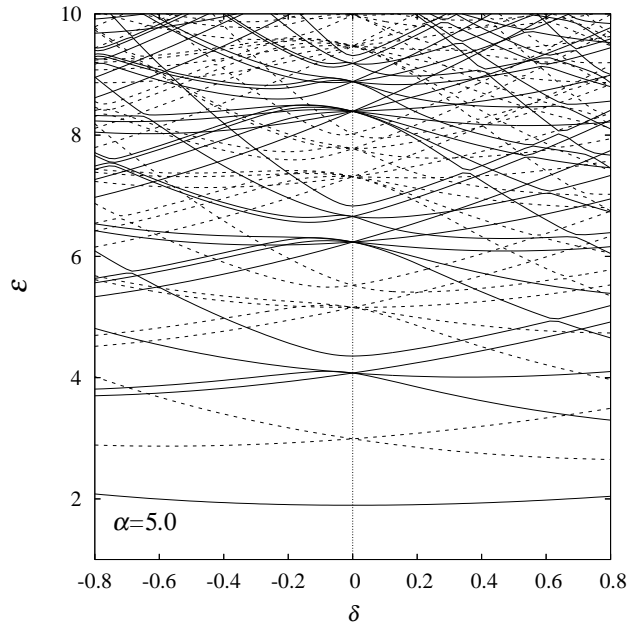}
\end{center}
\caption{\label{fig:nils5}
Single-particle level diagram for Hamiltonian (\ref{eq:hamiltonian})
with spheroidal deformation (\ref{eq:deform_sph}).  Scaled-energy
levels are plotted as functions of deformation parameter $\delta$
defined by Eq.~(\ref{eq:delta}).  Solid and broken lines represents
levels with even and odd parities, respectively.}
\end{figure}

Figure~\ref{fig:nils5} shows the single-particle spectra as functions
of spheroidal deformation parameter $\delta$.  The value of radial
parameter is put at $\alpha=5.0$, corresponding to medium-mass nuclei.
The degeneracies of levels at the spherical shape are resolved and shell
structure changes with varying deformation.  The level diagram is
similar to what is obtained for MO or WS/BP models without spin-orbit
coupling.  One of its characteristic features in comparison with the HO
model is the asymmetry of deformed shell structures in prolate and
oblate sides.  This asymmetry becomes more pronounced for larger
$\alpha$, and it might be regarded as the origin of prolate-shape
dominance in nuclear ground-state deformations.  We shall discuss the
semiclassical origin of the above asymmetry in the following
subsections.

\begin{figure}
\includegraphics[width=\linewidth]{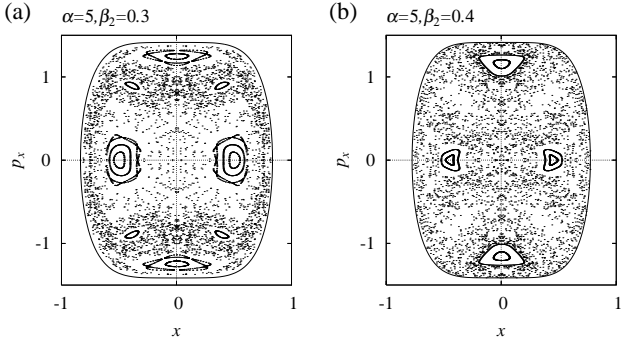}
\caption{\label{fig:pmap_beta}
Same as Fig.~\ref{fig:poincare} but for quadrupole
deformations $\beta_2=0.3$ and $0.4$ with $\alpha=5.0$.}
\end{figure}

In order to see the dependence on shape parametrization, we also
calculated the deformed quantum spectra for quadrupole deformation,
which might be more popular in earlier studies:
\begin{equation}
f(\theta)=\frac{1+\beta_2P_2(\cos\theta)}{\sqrt[3]{
 1+\frac35\beta_2^2+\frac{2}{35}\beta_2^3}}.
\end{equation}
The factor in the denominator arranges the conservation of volume
surrounded by equipotential surface.  Figure~\ref{fig:pmap_beta} show
Poincar\'{e} surface of section for quadrupole deformations
$\beta_2=0.3$ and $0.4$ with $\alpha=5.0$.  Comparing with the
Fig.~\ref{fig:poincare}(b), one will see that the particle
motions in the quadrupole potential are more chaotic than those in
the spheroidal potential.

\begin{figure}[b]
\begin{center}
\includegraphics[width=\linewidth]{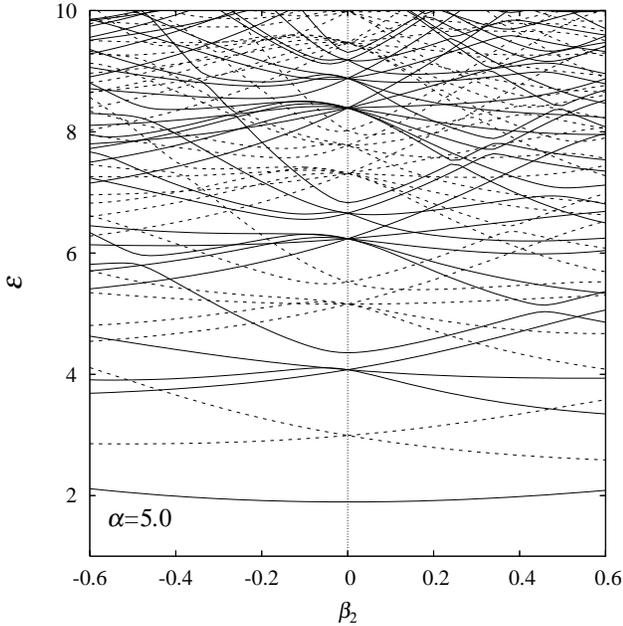}
\end{center}
\caption{\label{fig:nils5b}
Same as Fig.~\ref{fig:nils5} but as functions of quadrupole
deformation parameter $\beta_2$.}
\end{figure}

Figure~\ref{fig:nils5b} shows the level diagram for quadrupole
deformation.  Although the properties of classical motion are quite
different from those in spheroidal potential, the deformed shell
structures are very similar to each other.  Thus, the above difference
of shape parametrization does not cause a serious difference in the
gross shell structures at normal deformations.  Notable effects of
chaoticity in the quadrupole potential can only be seen in the strong level
repulsions at large deformations $\beta_2\gtrsim 0.3$.  Therefore, we
shall only consider the spheroidal deformation in the following
analysis.

\subsection{Prolate-oblate asymmetry in deformed shell structures}
\label{sect:pr-ob-asym}

Let us examine the properties of deformed shell structures in the
normal deformation region ($|\delta|\lesssim 0.3$).  As shown in
Figs.~\ref{fig:nils5} and \ref{fig:nils5b}, single-particle spectra in
a potential with a sharp surface show prolate-oblate asymmetry (in the
sense discussed in Sec.~I).  Hamamoto and Mottelson\cite{HamMot} paid
attention to the different ways of level \textit{fanning} (from the
terminology used in Ref.~\cite{HamMot}) in oblate and prolate sides;
level fanning is considerably suppressed in the oblate side as
compared to the prolate side.  Due to that suppression of level
fanning, shell structures in the oblate shapes are similar to those of
the spherical shape, and the system has a smaller chance to gain shell
energy by means of oblate deformation.  This may explain the feature
of prolate-shape dominance.  They have shown that the above asymmetric
way of level fanning can be understood from the interaction between
single-particle levels, which acts to suppress the level repulsions in
the oblate side for a potential with sharp surface.  It clearly explains
the fact that the asymmetry becomes more pronounced for heavier nuclei,
e.g., in the Woods-Saxon model\cite{StrMag}.  The same kind of asymmetry
is also found in the spectrum of the Nilsson model.

\begin{figure}
\begin{center}
\includegraphics[width=.9\linewidth]{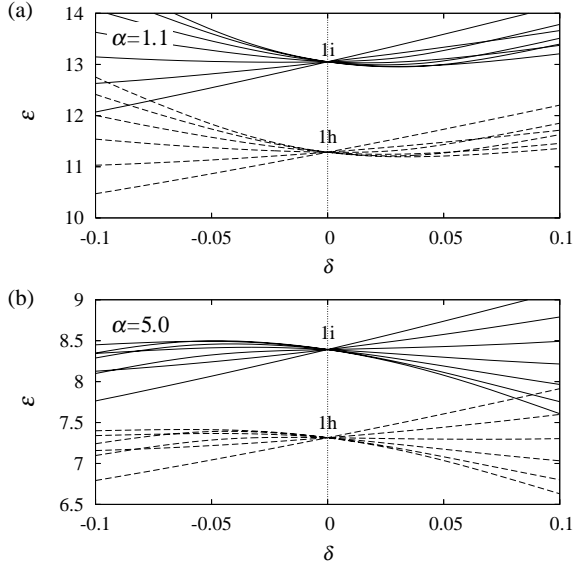}
\end{center}
\caption{\label{fig:nils_hi}
Fanning of single-particle levels 1h (dashed lines) and 1i (solid
lines) for (a) $\alpha=1.1$ and (b) 5.0.}
\end{figure}

In the spheroidal power-law potential model, the asymmetry in level
fanning becomes more pronounced for larger $\alpha$ as expected.  In
Fig.~\ref{fig:nils_hi}(b), fannings of some $nl$
levels ($n$ and $l$ represent principal and azimuthal quantum numbers,
respectively) are illustrated for $\alpha=5.0$.  One sees that the
level fannings are considerably suppressed in the oblate side as in
the cavity potential.

It is interesting to note that, if we take the radial parameter
$\alpha<2$ (although it does not correspond to actual nuclear
situations), the way of level fanning becomes just opposite to the
case of $\alpha>2$.  As one sees in
Fig.~\ref{fig:nils_hi}(a), level spreading is suppressed in the
prolate side.
We will discuss later if it causes oblate-shape dominance.

Following the analysis in Ref.~\cite{HamMot}, we calculate the
deformation energy
\begin{equation}
E_{\rm def}(A,\delta) = E(A,\delta)-E(A,0) \label{eq:edef1}
\end{equation}
and compare the energies in prolate and oblate sides at each local
minima.  Here, we assume the same single-particle spectra for neutrons
and protons and only consider $N=Z$ even-even nuclei for simplicity.
The sum of single-particle energies for nucleus of mass number $A$ is
given by
\begin{equation}
E_{\rm sp}(A)=4\sum_{i=1}^{n}e_i, \quad A=N+Z=4n.
\end{equation}
Using the Strutinsky method, the above energy can be decomposed into
a smooth part $\tilde{E}_{\rm sp}(A)$ and an oscillating part $\delta E(A)$.
As in usual, we can expect that the above oscillating part represents
the correct quantum shell effect of a many-body system.  In Strutinsky's
shell correction method, the smooth part is replaced with the
phenomenological liquid drop model (LDM) energy to get the total
many-body energy, but here we try to extract the smooth part also from
the single-particle energies.  In mean-field approximation, the single
particle Hamiltonian is written as
\begin{equation}
\hat{h}=\hat{t}+\hat{u},
\end{equation}
where $\hat{t}$ and $\hat{u}$ represent kinetic energy and mean-field
potential, respectively, and $\hat{u}$ is currently given by the
power-law potential.  In this case, by the use of the Virial theorem,
the average of $\hat{t}$ and $\hat{u}$ are in the ratio
$2\<\hat{t}\>=\alpha\<\hat{u}\>$, and one obtains
\begin{equation}
\<\hat{t}\>=\frac{\alpha}{\alpha+2}\<\hat{h}\>, \quad
\<\hat{u}\>=\frac{2}{\alpha+2}\<\hat{h}\>.
\end{equation}
Therefore, the smooth (average) part of the $A$-body energy is given
approximately by
\begin{equation}
\tilde{E}(A)\approx
\left\< \sum_i\hat{t}_i + \frac12\sum_i\hat{u}_i\right\>
=\frac{\alpha+1}{\alpha+2}\,\tilde{E}_{\rm sp}(A) \label{eq:etildeofA}
\end{equation}
This expression will be valid for many-body systems interacting with
two-body interaction.  Thus, we evaluate the $A$-body energy by
\begin{equation}
E(A)=\frac{\alpha+1}{\alpha+2}\tilde{E}_{\rm sp}(A)+\delta E(A)
\label{eq:eofA}
\end{equation}

\begin{figure}
\begin{center}
\ifonline
\includegraphics[width=\linewidth]{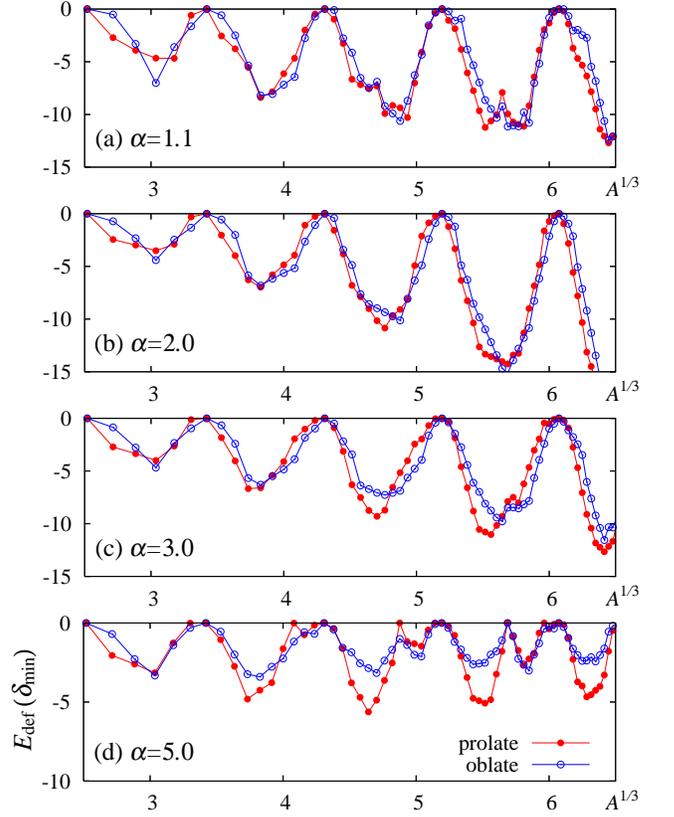}
\else
\includegraphics[width=\linewidth]{Fig14.ps}
\fi
\end{center}
\caption{\label{fig:emin}
(Color online)
Deformation energies (\ref{eq:edef1}) of prolate and oblate states at
each local minima.  Deformation energy $E_A(\delta)$ is
calculated as a function of deformation $\delta$ and its minimum values
for prolate and oblate sides are plotted with filled and open circles,
respectively.}
\end{figure}

Figure~\ref{fig:emin} compares the local minima of deformation
energies (\ref{eq:edef1}) in prolate and oblate sides.  At the HO
value [$\alpha=2.0$, panel (b)], prolate and oblate deformed shell
structures are symmetric and the deformation energies are comparable
with each other.  For $\alpha>2$ [panels (c) and (d) of
Fig.~\ref{fig:emin}], the deformation energies in the prolate side become
considerably lower than in the oblate side as the radial parameter
$\alpha$ becomes larger.  The power-law potential model thus reproduce
correctly the feature of prolate-shape dominance in nuclear
deformation.

For $\alpha<2$, as shown in Fig.~\ref{fig:emin}(a), we
find no indication of oblate-shape dominance in spite of the feature
of level fanning shown in Fig.~\ref{fig:nils_hi}(a).
One finds some lowest energy states at oblate shapes $\delta\sim -0.3$,
but the difference in energies between prolate and oblate minima are
generally small.  Therefore, one cannot fully explain the
prolate(oblate)-shape dominance only by the ways of level fanning.

\begin{figure}
\begin{center}
\ifonline
\includegraphics[width=\linewidth]{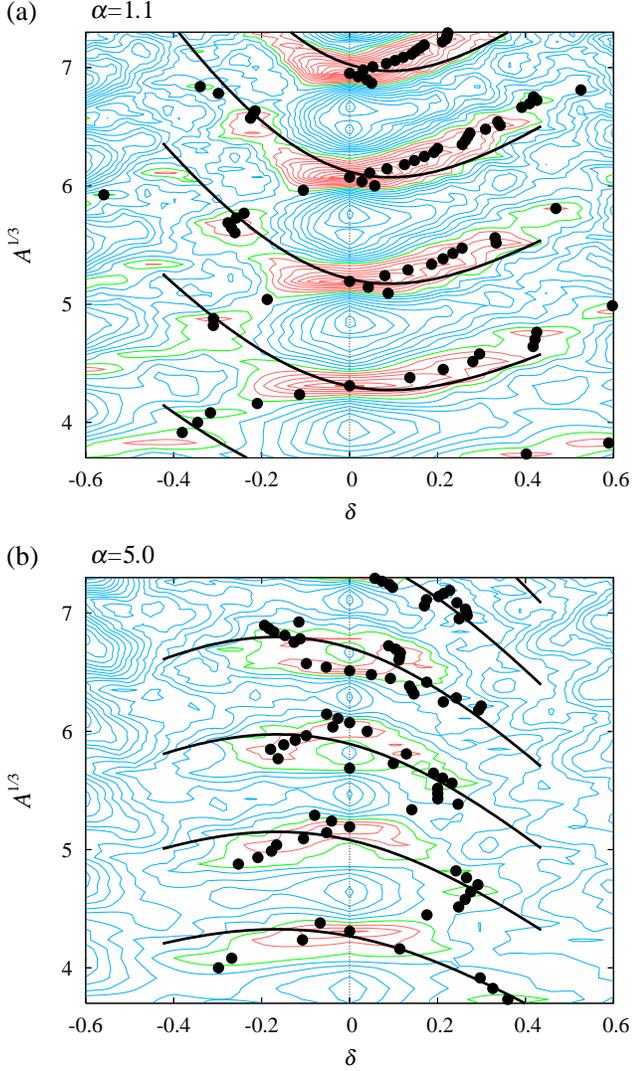}
\else
\includegraphics[width=\linewidth]{Fig15.ps}
\fi
\end{center}
\caption{\label{fig:sce}
(Color online)
Contour plot of the $A$-body shell-deformation energy
(\ref{eq:edef2}) in the deformation-mass number plane
$(\delta,A^{1/3})$ for (a) $\alpha=1.1$ and (b) 5.0.
Solid and dashed contour lines represent negative and
positive values, respectively.  Dots represent values of
the deformation parameter at absolute energy minima for each $A$.
Thick solid curves represent constant-action lines
(\ref{eq:ca-line_sce}) for bridge orbit M(1,1).}
\end{figure}

In order to analyze shape stability, we define
\textit{shell-deformation energy} using the smooth part of the energy at
spherical shape as a reference,
\begin{equation}
\Delta E(A,\delta)=E(A,\delta)-\tilde{E}(A,0), \label{eq:edef2}
\end{equation}
with Eqs.~(\ref{eq:etildeofA}) and (\ref{eq:eofA}).  [Note that the
second term on the right-hand side of Eq.~(\ref{eq:edef2}) is not
$\tilde{E}(A,\delta)$, so that $\Delta E$ contains the smooth part of
the deformation energy.]  Figure~\ref{fig:sce} shows contour plots of
$\Delta E$ for $\alpha=1.1$ and $5.0$ as functions of deformation
$\delta$ and mass number $A$.  They show some deep minima along the
$\delta=0$ axis at values of $A$ corresponding to spherical magic
numbers.  The energy valleys run through these minima and the
deformation energy minima distribute along them.  For $\alpha=5.0$,
the valley lines in the $(\delta,\tau)$ plane have large slopes in the
prolate side and deep energy minima are formed around $\delta\sim 0.2$
for mass numbers at the middle of adjacent spherical magic numbers,
while the valley lines are almost flat in the oblate side.  This is
essentially the same behavior as what Frisk found for the spheroidal
cavity\cite{Frisk}.  For $\alpha=1.1$, the valley lines have larger
slope in the oblate side, but the slope in the prolate side is not as
small as in the oblate side for $\alpha=5.0$ and the deformation
energy minima distribute mainly along the valley lines in the prolate
side.  One can find rather deep energy minima at $\delta\sim -0.3$ for
particle numbers between the spherical magic numbers, but the energy
difference between oblate and prolate local minima are generally
small.  Thus, for an understanding of prolate-shape dominance, it is
critical to explain the asymmetric behavior of the slopes of the
energy valleys.

\begin{figure}
\begin{center}
\ifonline
\includegraphics[width=\linewidth]{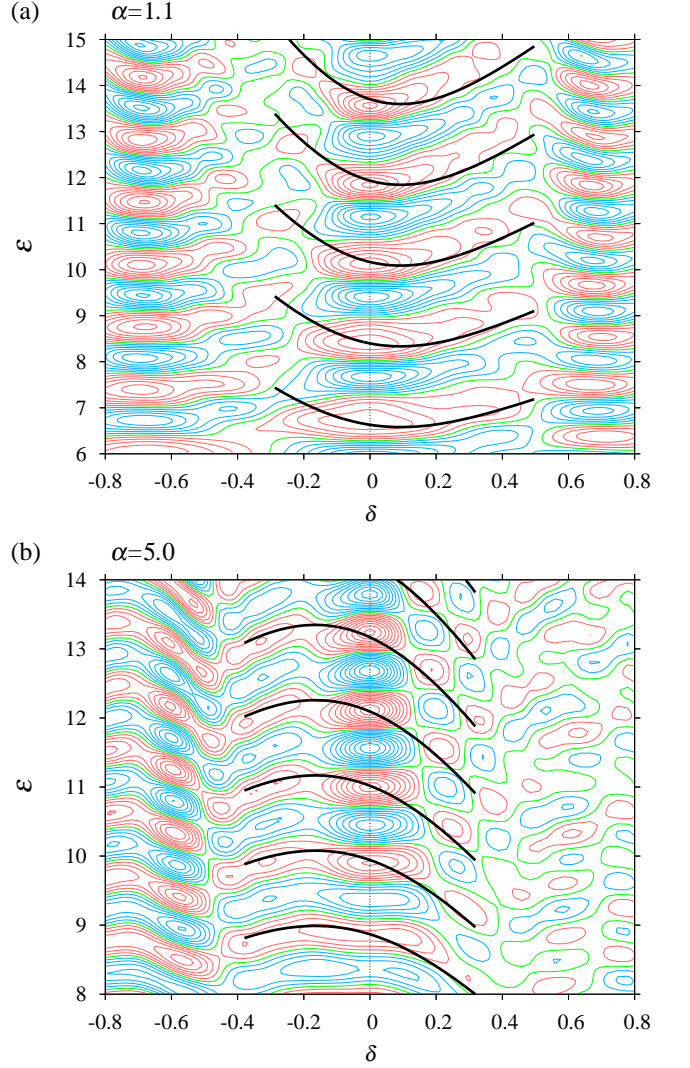}
\else
\includegraphics[width=\linewidth]{Fig16.ps}
\fi
\end{center}
\caption{\label{fig:sld}
(Color online) Contour plot of oscillating level density for radial
parameter (a) $\alpha=1.1$ and (b) 5.0 in deformation-energy plane
$(\delta,\cE)$.  For calculation of coarse-grained level density,
smoothing width $\Delta\cE=3.0$ is taken.  Solid and dashed contour
lines represent negative and positive values, respectively.  Thick
solid lines represent constant-action lines (\ref{eq:ca-line}) for the
bridge orbits M(1,1).}
\end{figure}

Since shell energy takes a deep negative value when the
single-particle level density at the Fermi energy is low, let us
investigate the coarse-grained single-particle level density as
functions of energy and deformation.  Figure~\ref{fig:sld} shows the
oscillating part of course-grained single-particle level density for
$\alpha=1.1$ and $5.0$ plotted as functions of deformation $\delta$
and scaled energy $\cE$.  They show regular ridge-valley structures
similar to the shell energy.  Therefore, for an understanding of
prolate-shape dominance, it is essential to investigate the origin of
the above ridge-valley structures in the deformed single-particle
level density.  For the spheroidal cavity potential, Frisk ascribed it
to the change of the action integrals along the triangular and
rhomboidal orbits in the meridian plane, for which the volume
conservation condition plays an important role\cite{Frisk}.  We are
going to study the case of the more realistic power-law potential.

In the following subsections, we will show that the above ridge-valley
structure can be explained in connection with classical periodic
orbits using semiclassical periodic-orbit theory.  The thick solid
curves in Figs.~\ref{fig:sce} and \ref{fig:sld} represent the
semiclassical prediction of the valley lines which will be discussed
in Sec.~\ref{sec:pd_semi}.

\subsection{Periodic orbits in spheroidal potential}
\label{sec:po_spheroid}

In order to examine the semiclassical origin of the above asymmetry in
a deformed shell structure using periodic orbit theory, we first
consider the properties of classical periodic orbits in the spheroidal
power-law potential and their bifurcations.  For the spherical
potential, all the periodic orbits are planar and degenerate with
respect to rotations.  The degree of degeneracy for the orbit family
is described by degeneracy parameter $\cK$ which represents the number
of independent continuous parameters required to specify a certain
orbit in the family.  The maximum value of $\cK$ is equal to the
number of independent symmetric transformations of the system.  The
isolated orbits have $\cK=0$.  In the spherical cavity potential,
degeneracy parameter is $\cK=3$ for generic periodic orbits, and
$\cK=2$ for diametric and circular orbits which are transformed onto
themselves by one of the rotations.  If the spheroidal deformation is
added to the potential, generic planar orbits bifurcate into two
branches: One is the orbit in the equatorial plane and the other is
the orbit in the meridian plane (the plane containing the symmetry
axis).  All but two exceptional orbits degenerate with respect to the
rotation about the symmetry axis, and the degeneracy parameter is
$\cK=1$.  The diametric orbit bifurcate into degenerate family of
equatorial diametric orbits ($\cK=1$) and an isolated diametric orbit
along the symmetry axis ($\cK=0$).  The circular orbit bifurcates into
an isolated equatorial circular orbit ($\cK=0$) and an oval-shape
orbit in the meridian plane ($\cK=1$).  With increasing deformation
towards prolate side $(\delta>0)$, the equatorial orbits undergo
successive period $m$-upling bifurcations and new 3D orbits emerge.  In
the oblate side, the diametric orbit along the symmetry axis undergoes
successive period $m$-upling bifurcations and generates new
meridian-plane orbits.  These new-born 3D and meridian-plane orbits
have hyperbolic caustics and are sometimes called \textit{hyperbolic
orbits}.

It is very interesting to note that the above new-born hyperbolic
orbits from equatorial orbits are distorted towards the symmetry axis
with increasing deformation and finally submerge into the diametric
orbit along the symmetry axis.  (Some 3D orbits submerge into other
hyperbolic orbits before submerging into the symmetry-axis orbit at
last.)  In this way, the hyperbolic orbits make bridges between the
equatorial and symmetry-axis orbits, and we shall call those
hyperbolic orbits ``bridge orbits''\cite{ABbridge,BarDav87}.  With
increasing $\delta$, periods of the equatorial orbit decrease while
that of the symmetry-axis orbit increases.  At each crossing point of
the periods (or actions) of repeated equatorial and symmetry-axis
orbits, bridge orbits exist to intervene between them.

Accordingly, we shall classify periodic orbits in the spheroidal
power-law potential into the following four groups:
\begin{enumerate}\def\labelenumi{\roman{enumi})}
\item \textit{Isolated orbits} ($\cK=0$):  This group consists of the
diametric orbit along the symmetry axis ($z$-axis), denoted Z, and the
circular orbit in the equatorial plane, denoted EC.  Orbit EC is
stable both in the prolate and oblate sides, whose repeated
period $m$-upling bifurcations generate 3D bridge orbits.  Orbit Z is
stable in the oblate side and undergoes successive period $m$-upling
bifurcations, while its stability alternates in the prolate side with
repeated bifurcations which absorb bridge orbits.
\item \textit{Equatorial-plane orbits} ($\cK=1$):  This corresponds to
the equatorial-plane branch of the deformation-induced bifurcation.
They have the same shapes as those in the spherical potential shown in
Fig.~\ref{fig:orbit_sph}.  They are denoted E$(k,m)$, where $k$ is the
number of vertices (corners) and $m$ is the number of rotation.
The diametric orbit is specially denoted X (which includes the orbits
along the $x$ axis).
\item \textit{Meridian-plane orbits} ($\cK=1$):  This corresponds to
the meridian-plane branch of the deformation-induced bifurcation.
They survive up to any large deformation, keeping their original
geometries.
\item \textit{Bridge orbits} ($\cK=1$):  These orbits emerge from the
bifurcations of equatorial orbits.  Meridian-plane orbits emerge from
diametric orbits and submerge into repetitions of orbit Z.  Nonplanar
3D orbits emerge from nondiametric equatorial orbits, and they also
submerge into the orbit Z.  Some of them submerge into other bridge
orbits before submerging into Z.  The meridian-plane bridge which rests
between $m$X ($m$th repetition of X) and $n$Z ($n$th repetition of
Z) is denoted M$(m,n)$.  Except for the M(1,1) bridge, a pair of stable
and unstable bridge orbits emerge, and are denoted M$(m,n)_s$ and
M$(m,n)_u$, respectively.  3D bridge B$(m,m,n)$ emerges via
pitchfork bifurcation of equatorial $m$EC ($m$th repetitions
of EC), and submerges into M$(m,n)$ orbit before finally submerging into
$n$Z.  The other 3D bridges intervening between equatorial E$(k,m)$ and $n$Z
emerge as a stable and unstable pair, and are denoted as
B$(k,m,n)_{s,u}$.  With increasing deformation, they first submerge into
3D bridge B$(m,m,n)$, which will submerge into M$(m,n)$ and finally
into $n$Z.
\end{enumerate}

\begin{figure}
\begin{center}
\includegraphics[width=.8\linewidth]{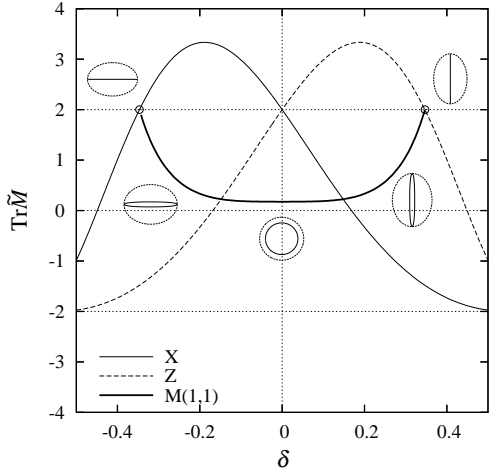}
\end{center}
\caption{\label{fig:bif_diameter1}
Bifurcation diagram for the M(1,1) bridge orbit between the X and Z
orbits for $\alpha=3.0$.  Traces of the symmetry-reduced monodromy
matrix $\tilde{\BM}$ are plotted as functions of deformation parameter
$\delta$.  Bifurcation points ($\Tr\tilde{\BM}=2$) are indicated by
open circles.  Shapes of the periodic orbits as well as the
equipotential surface are also shown.}
\end{figure}

Figure~\ref{fig:bif_diameter1} shows the bifurcation diagram for bridge
orbit M(1,1) for $\alpha=3.0$.  The traces of $(2\times 2)$
symmetry-reduced monodromy matrices for relevant periodic orbits are
plotted as functions of deformation parameter $\delta$.  The $\cK=1$
family of equatorial diameter orbits X undergoes pitchfork bifurcation
at $\delta=-0.34$ and a family of oval-shape meridian-plane orbits
M(1,1) emerge.  In the limit $\delta\to 0$, the shape of the M(1,1) orbit
approaches a circle, and it associates with equatorial circular orbit EC
to form a $\cK=2$ family.  At $\delta>0$ it bifurcates into an equatorial
EC and a meridian M(1,1) family again.  The meridian branch submerges
into the orbit Z at $\delta=0.34$ via pitchfork bifurcation.  Thus,
the orbits M(1,1) make a bridge between two diametric orbits X and Z.

In the HO limit, $\alpha\to 2$, the bridge shrinks to a crossing point
of X and Z orbits and can exist only at $\delta=0$ (spherical shape),
where they altogether form a degenerate $\cK=2$ family.
With increasing $\alpha$, the bridge orbit exists in a wider range of
deformation over the crossing point.
In the cavity limit, $\alpha\to\infty$, this orbit
approaches the so-called \textit{creeping orbit} or \textit{whispering
gallery orbit}, which runs along the boundary.

\subsection{Semiclassical origin of prolate-oblate asymmetry}
\label{sec:pd_semi}

\begin{figure}
\begin{center}
\ifonline
\includegraphics[width=\linewidth]{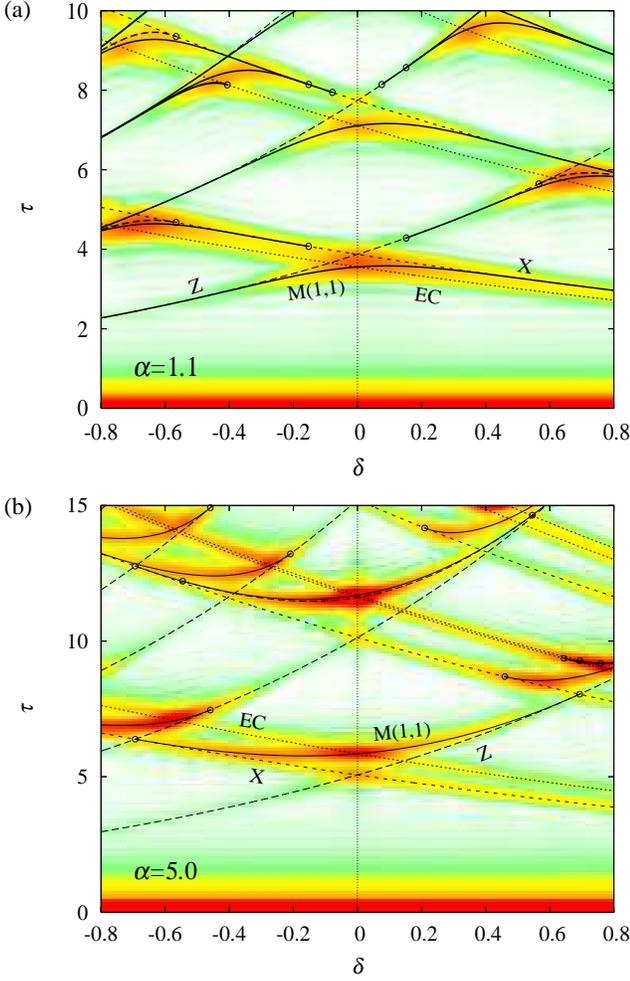}
\else
\includegraphics[width=\linewidth]{Fig18.ps}
\fi
\end{center}
\caption{\label{fig:fourier}
(Color online)
Modulus of the Fourier transform of the level density is shown by
gray-scale plot as a function of deformation parameter $\delta$ and
scaled period $\tau$.  Scaled periods of classical periodic orbits are
also displayed.}
\end{figure}

To see the effect of the above bifurcation on the shell structure, we
calculate Fourier transform of level density (\ref{eq:fourier_qm})
with the obtained quantum spectra.  In Fig.~\ref{fig:fourier}, modulus
of Fourier transform $|F(\tau,\delta)|$ is shown in a gray-scale plot
as a function of deformation $\delta$ and scaled period $\tau$.
Scaled periods of classical periodic orbits $\tau_\beta(\delta)$ are
also drawn by lines.  One sees a nice correspondence between the
quantum Fourier amplitude and classical periodic orbits.
Particularly, one can find significant peaks along the bridge orbit
M(1,1), which indicate that the shell structure in the normal deformation
region is mainly determined by the contribution of this bridge orbit.

Let us assume that a contribution of single orbit (or degenerate family)
$\beta$ dominates the periodic orbit sum, namely,
\begin{equation}
\delta g(\cE) \approx A_\beta \cos(\cE \tau_\beta-\tfrac{\pi}{2}\nu_\beta).
\end{equation}
Then, the valley lines of level density should run along the curves
where the above cosine function takes the minimum value $-1$, namely,
\begin{equation}
\cE\tau_\beta-\frac{\pi}{2}\nu_\beta = (2n+1)\pi, \quad n=0,1,2,\cdots.
\end{equation}
In Fig.~\ref{fig:sld}, we plot these constant-action lines
\begin{equation}
\cE = \frac{(2n+1+\frac12\nu_\beta)\pi}{\tau_\beta(\delta)}
\label{eq:ca-line}
\end{equation}
for bridge orbit M(1,1).  We see that the constant-action lines of the
bridge orbit nicely explain the ridge-valley structure in the quantum
level density.
The slight disagreement of constant-action lines and the bottom of
the energy valleys might be due to interference between other PO
contributions.

The shell energy is also given by the periodic-orbit contribution as
\begin{equation}
\delta\cE(N)\approx\frac{A_\beta}{\tau_\beta^2}
 \cos[\cE_F(N)\tau_\beta-\tfrac{\pi}{2}\nu_\beta],
\end{equation}
where $\cE_F$ represents Fermi level, which is approximately given by
\begin{equation}
\cE_F(N)\approx\left[3N/c_0(\alpha)\right]^{1/3}
\end{equation}
which is derived from the leading term of Eq.~(\ref{eq:nbar_etf}).
Thus, the shell energy takes large negative values along the
constant-action lines for dominant orbit $\beta$:
\begin{equation}
\cE_F\approx\left(\frac{3}{c_0(\alpha)}\,\frac{A}{4}\right)^{1/3}
=\frac{(2n+1+\tfrac12\nu_\beta)\pi}{\tau_\beta(\delta)}.
\label{eq:ca-line_sce}
\end{equation}
In Fig.~\ref{fig:sce}, we also plot the above constant-action lines
for bridge orbits M(1,1) with thick solid lines.  They satisfactorily
explain the valley lines of shell energy.
Distribution of deformed shell energy minima in
Fig.~\ref{fig:sce} are thus understood as the effect of bridge orbit
contribution.

For $\alpha>2$, bridge orbits appear upward from the crossing point of
two diametric orbits X and Z in the $(\delta,\tau)$ plane.  Note that the
scaled action of orbit Z has a larger slope than that of orbit X in the
$(\delta,\tau)$ plane.  This difference comes from the fact that the
lengths of semiaxes $R_z$ and $R_\perp$ in a volume-conserved
spheroidal body are proportional to the different powers of
deformation parameter $\delta$ as in Eq.~(\ref{eq:delta}).  The scaled
period of the diametric orbit along the $i$th axis is proportional to the
length of corresponding semiaxis $R_i$,
\[
\tau_i=\tau_0^D\frac{R_i}{R_0},
\]
where $\tau_0^D$ is the scaled period of the diametric orbit at
spherical shape.  Using Eq.~(\ref{eq:delta}), one has
\begin{equation}
\tau_X=\tau_0^D e^{-\frac13\delta},\quad
\tau_Z=\tau_0^D e^{\frac23\delta}. \label{eq:tauxz}
\end{equation}
Therefore, the bridge between X and Z orbits has a large slope in the
prolate side while it is almost flat in the oblate side.  This clearly
explains the profile of ridge-valley structures in level density and
shell energy.  With increasing $\alpha$, triangular- and square-type
orbits emerge at $\alpha=7$ and $14$, respectively, via the
\textit{isochronous} bifurcations of the circular orbit [see
Eq.~(\ref{eq:bifpts}) for $m=1$] for spherical shape.  With spheroidal
deformation, they bifurcate into equatorial and meridian branches,
which are both singly degenerated due to the axial symmetry.  For
finite $\alpha$, they submerge into oval orbits and finally into
diametric orbits at large deformations.  In this sense, they are also
bridge orbits intervening between two diametric orbits.  These
meridian orbits survive up to larger deformation with increasing
$\alpha$, and in the cavity limit ($\alpha\to\infty$), they survive
for any large deformation.  Therefore, the meridian orbits in the cavity
potential can be regarded as a limit of bridge orbits.  Thus we see
that Frisk's argument for a spheroidal cavity system\cite{Frisk} is
continuously extended to the case of finite diffuseness.

For $\alpha<2$, a bridge orbit appear in the opposite side of the crossing
point and its slope becomes larger in the oblate side.  This also explains
the profile of valley lines in level density and shell energy for
$\alpha=1.1$ as shown in Figs.~\ref{fig:sld}(a) and \ref{fig:sce}(a).

\subsection{Superdeformed shell structures}

In the axially deformed harmonic oscillator (HO) potential model, one
sees simultaneous degeneracy of many energy levels at rational axis
ratios.  The HO model is often used for the nuclear mean-field
potential in the limit of light nuclei.  In the HO model,
superdeformation is explained as the result of strong level
bunching at axis ratio 2:1.  A search for much larger deformation
originated from the strong level bunching at axis ratio 3:1 (sometimes
referred to as \textit{hyperdeformation}) has also been a challenging
experimental and theoretical problem.  On the other hand, the
spheroidal cavity model, which is used as the limit of potential for
heavy nuclei, also shows superdeformed shell structures, while the
shell effect is much weaker than that found in the oscillator model.
In the spheroidal cavity model, superdeformed shell structures are
intimately related with emergence of meridian and 3D orbits which
oscillate twice in the short axis direction while they oscillate once
in the long-axis direction, just as the degenerate 3D orbits in the 2:1
axially-deformed HO potential\cite{ASM}.  One may expect to have a
\textit{unified} semiclassical understanding of the origin of
superdeformed shell structures found in the above two limiting cases
by connecting them with the power-law potential model.

\begin{figure}
\ifonline
\includegraphics[width=\linewidth]{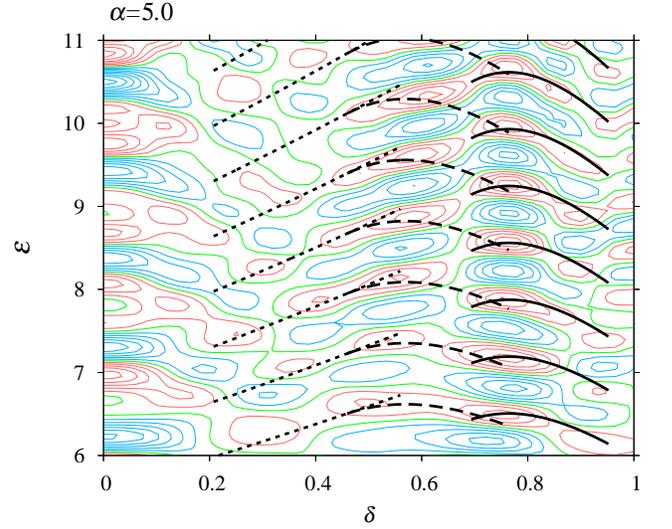}
\else
\includegraphics[width=\linewidth]{Fig19.ps}
\fi
\caption{\label{fig:sld_sd}
(Color online)
Contour plot of oscillating level density for radial parameter
$\alpha=5.0$ around a prolate superdeformed region.  
Smoothing width $\varGamma=0.2$ is used.
Solid and dashed contour lines represent negative and positive values,
respectively.
Thick dotted, dashed and solid lines represent constant action lines
(\ref{eq:ca-line}) of periodic orbits 2X, M(2,1), and B(2,2,1),
respectively.}
\end{figure}

Figure~\ref{fig:sld_sd} shows the oscillating part of the
coarse-grained level density for radial parameter $\alpha=5.0$ and
deformation $\delta$ around a superdeformed region.  It clearly show
that new regularities in shell structure are formed at a superdeformed
region.  The valley lines are up-going till $\delta\sim 0.5$, and
they bend down around $\delta\sim 0.6$.  One sees another deep minima at
$\delta\gtrsim 0.7$.  Let us examine their semiclassical origins.

\begin{figure}[p]
\begin{center}
\includegraphics[width=.85\linewidth]{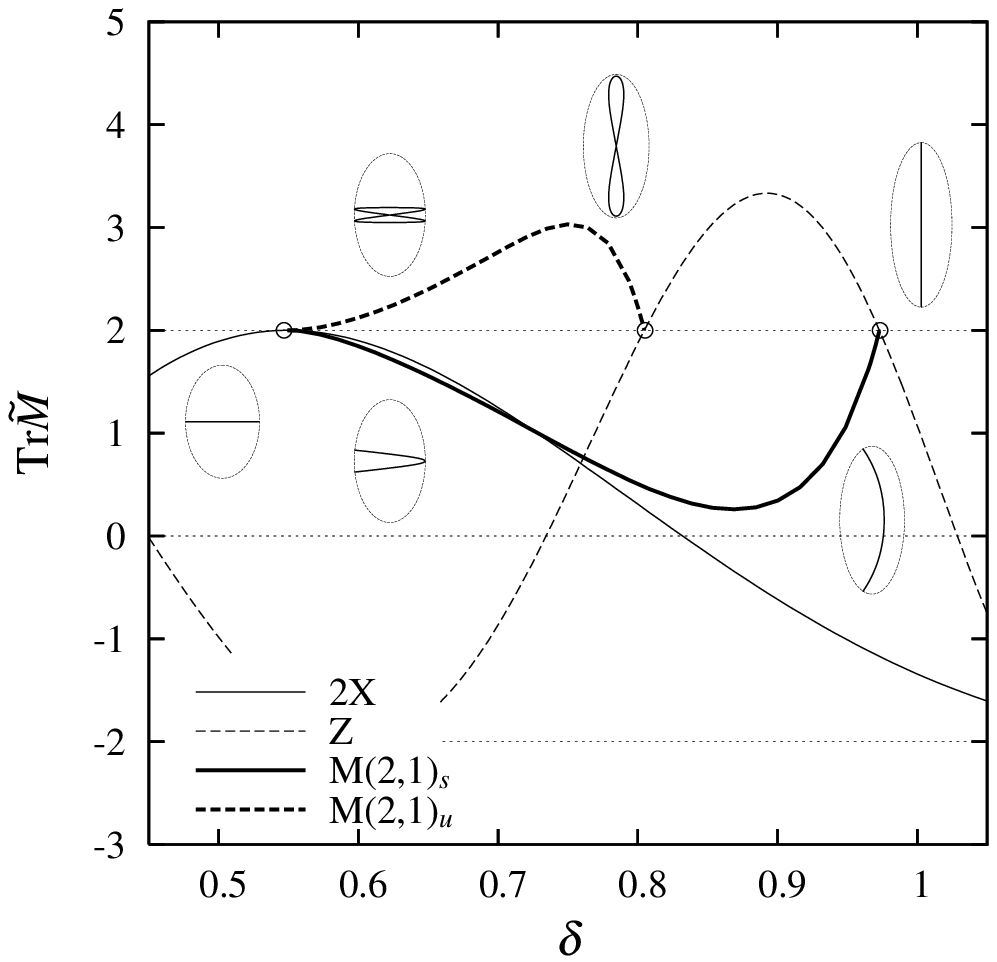}
\end{center}
\caption{\label{fig:bif_diameter2}
Same as Fig.~\ref{fig:bif_diameter1} but for the M(2,1) bridge orbit
between 2X and Z; ``2X'' represents the second repetition of X.  The
radial parameter $\alpha=3.0$ is used.  A pair of bridge orbits emerge
at $\delta=0.55$ via period-doubling bifurcation.  The unstable branch
M$(2,1)_u$ and stable branch M$(2,1)_s$ submerge into Z at
$\delta=0.80$ and 0.97, respectively, via pitchfork bifurcations.}
\bigskip

\begin{center}
\includegraphics[width=\linewidth,bb=0 288 510 576,clip]{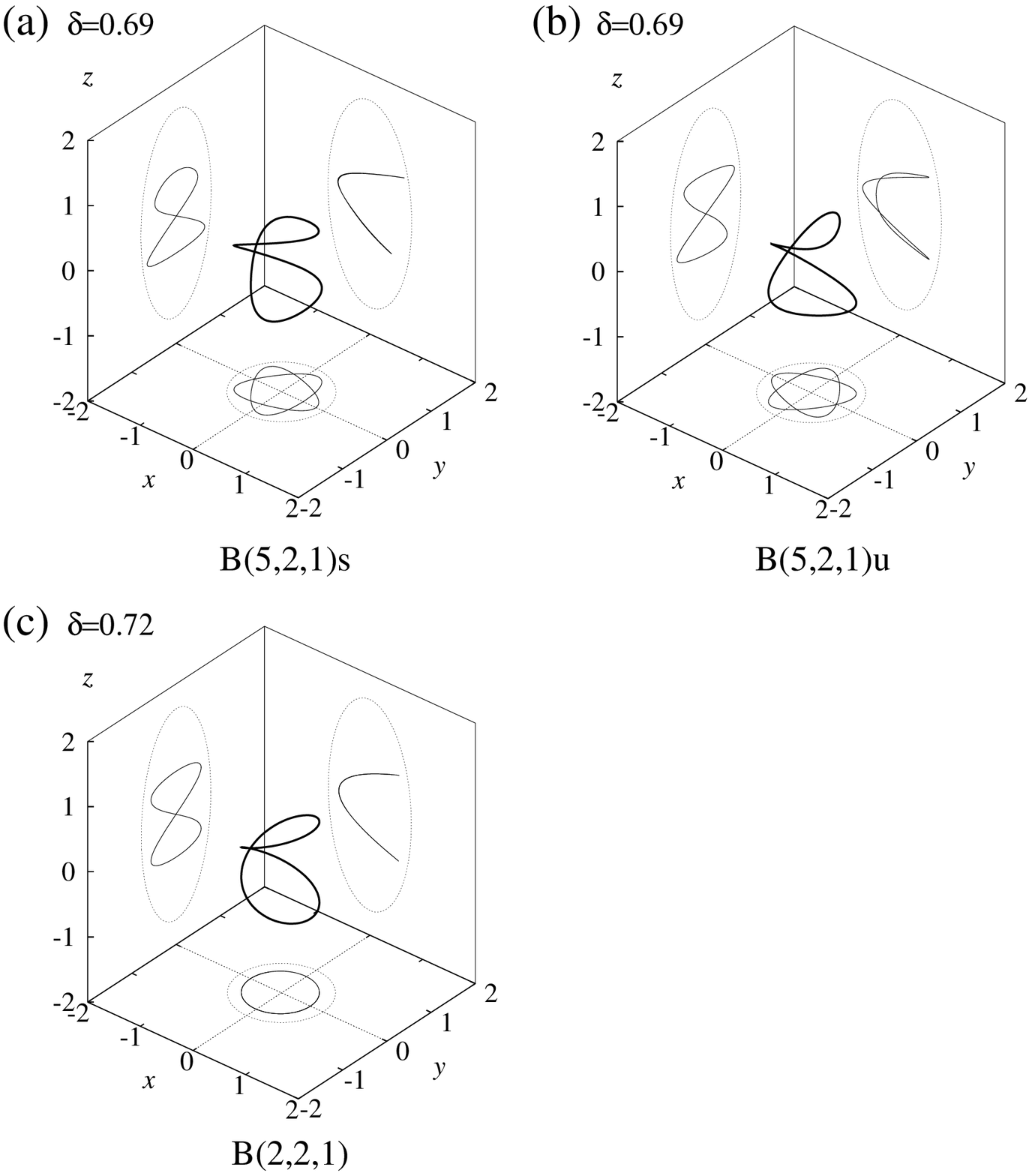}\\[5mm]
\includegraphics[width=.5\linewidth,bb=0 0 255 288,clip]{Fig21.ps}
\end{center}
\caption{\label{fig:orbit3d_sd}
3D orbits responsible for the superdeformed shell
structure at $\delta\sim 0.7$ and $\alpha=5$.
3D plots and projections on $(x,y)$, $(x,z)$, and $(y,z)$ planes are
shown as well as equipotential surfaces.
B$(5,2,1)_s$ and B$(5,2,1)_u$ are a
pair of stable and unstable 3D orbits that emerged from
equatorial orbit E(5,2).  B(2,2,1) emerged from the
second repetition of the equatorial circular orbit, 2EC.}
\end{figure}

For $\alpha>2$, one finds bridge orbits M(2,1) which intervenes
between orbits 2X (second repetitions of X) and Z.
Figure~\ref{fig:bif_diameter2} is the bifurcation diagram for the
orbits relevant to this bifurcation, calculated for $\alpha=3.0$.  The
orbit X undergoes a period-doubling bifurcation at $\delta=0.55$ and
there emerges a pair of bridge orbits M$(2,1)_s$ (stable) and
M$(2,1)_u$ (unstable).  They have shapes of a boomerang and a
butterfly as shown in Fig.~\ref{fig:bif_diameter2}.  With increasing
$\delta$, those orbits are distorted toward the $z$ axis and finally
submerge into the orbit Z at different values of $\delta$ via
pitchfork bifurcations.

For larger $\alpha$, various equatorial orbits appear as shown in
Fig.~\ref{fig:orbit_sph}, and they also undergo bifurcations by
imposing deformation.  Each of those bifurcations will generate a pair
of 3D bridge orbits, which are also distorted toward the symmetry axis
by increasing $\delta$ and finally submerge into Z.
Figure~\ref{fig:orbit3d_sd} shows some 3D bridge orbits important for
superdeformed shell structures for $\alpha=5.0$.  Equatorial circular
orbit EC undergoes a period-doubling bifurcation, which is peculiar to 3D
systems, and generates 3D bridge orbit B(2,2,1).  Equatorial orbit E(5,2)
undergoes a nongeneric period-doubling bifurcation and a pair of 3D
bridge orbits B$(5,2,1)_{s,u}$ emerge.  All the above 3D orbits finally
submerge into the Z orbit by increasing deformation $\delta$.  See
the Appendix for a detailed description of these 3D bridge
orbit bifurcations.

\begin{figure}
\ifonline
\includegraphics[width=\linewidth]{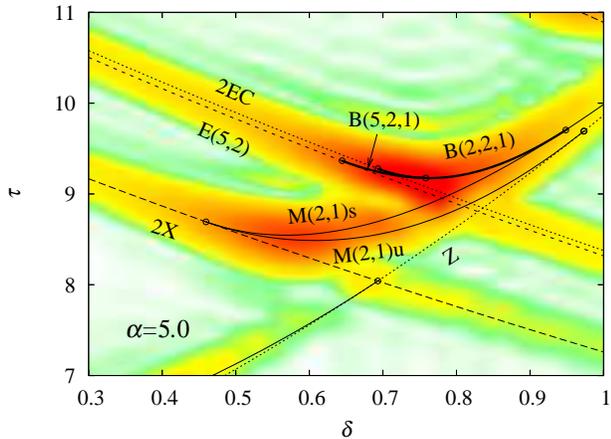}
\else
\includegraphics[width=\linewidth]{Fig22.ps}
\fi
\caption{\label{fig:ftlmap_sd}
(Color online)
Gray-scale plot of the Fourier transform of the quantum level density
(\ref{eq:fourier_qm}) for radial parameter $\alpha=5.0$ as a function of
deformation $\delta$ and scaled period $\tau$.  The modulus of the
Fourier transform has a large value in the dark region.  The scaled periods
of classical periodic orbits $\tau_\beta$ are displayed with lines.  Their
bifurcation points are indicated by open circles.}
\end{figure}

Figure~\ref{fig:ftlmap_sd} shows the Fourier transform of scaled-energy
level density for $\alpha=5.0$ around a superdeformed region.
The scaled periods of classical periodic orbits are also drawn with lines.
The Fourier amplitude shows remarkable enhancement along the bridge
orbits M(2,1) and B(5,2,1), indicating their significant roles in
superdeformed shell structures.

The constant action lines (\ref{eq:ca-line}) for M(2,1) and B(2,2,1)
are shown in Fig.~\ref{fig:sld_sd} with thick solid and broken
lines.  They perfectly explain the ridge-valley structures of quantum
level densities.  This shows the significant roles of bifurcations of
M(2,1) and B$(n,2,1)$ orbits for enhanced shell effects at $\delta\sim
0.5$ and 0.7, respectively.

M(2,1) and B(2,2,1) orbits shrink to the crossing point of 2X (2EC)
and Z orbits in the HO limit, $\alpha\to 2$, and turn into a $\cK=4$
degenerate family.  With increasing $\alpha$, the deformation range in
which a bridge orbit can exist becomes wider.  Therefore, the
bifurcation deformation of the orbits 2X and 2EC becomes smaller with
increasing $\alpha$, and the effect of these orbits takes place at
smaller deformation.  This may explain the experimental fact that
deformations of the superdeformed states are smaller for heavier
nuclei; e.g., $\beta_2\sim 0.6$ for the Dy region and $\beta_2=0.4 \mbox{--}
0.5$ for the Hg region\cite{Twin,AAberg}.  In the cavity limit
$\alpha\to\infty$, the two meridian orbits M(2,1) and the two 3D
orbits B($n$,2,1) respectively join to form $\cK=1$ families, which
survive for arbitrary large deformations.\cite{Hide,ASM}

In conclusion, the highly degenerate family of orbits in the rational
HO potential ($\alpha=2$) are \textit{resolved} at $\alpha>2$ into two
orbits: equatorial and symmetry-axis orbits that have fewer
degeneracies, and the bridge orbit which mediates between them within
a finite deformation range.  The ``length'' of the bridge in the
$(\delta,\tau)$ plane grows with radial parameter $\alpha$, and the
superdeformed shell structures are formed in smaller $\delta$ for
large $\alpha$, corresponding to heavier nuclei, due to the strong
shell effect brought about by the bridge-orbit bifurcation.  In the
$\alpha\to\infty$ limit, some of the simplest bridge orbits coincide
with meridian and 3D orbits emerging from the bifurcation of
equatorial orbits at $\delta=0.5 \mbox{--} 0.6$ which play a significant
role in superdeformed shell structures in the spheroidal cavity
model\cite{ASM,Magner2}.  Thus, the semiclassical origins of
superdeformed and hyperdeformed shell structures in the HO and the
cavity models are unified as the two limiting cases of the
contribution of bridge-orbit bifurcations in power-law potential
model.

\section{Summary}
\label{sect:summary}

We have made a semiclassical analysis of deformed shell structures
with the radial power-law potential model, which we introduce as a
realistic nuclear mean-field model (except for the lack of a
spin-orbit term in the current version) for stable nuclei in place of
WS/BP models.  We have shown that bridge orbits mediating equatorial
and symmetry-axis orbits play a significant role in normal and
superdeformed shell structures.  Particularly, prolate-oblate
asymmetry of deformed shell structures, which is responsible for the
prolate dominance in nuclear deformations, is clearly understood as
the asymmetric slopes of bridge orbits in the $(\delta,\tau)$ plane.
This asymmetry grows with increasing radial parameter $\alpha$, and
thus with increasing mass number $A$, which explains the fact that
the prolate dominance is more remarkable in heavier nuclei.  Some of
these bridge orbits coincide with triangular and rhomboidal orbits in
the cavity limit $\alpha=\infty$, whose significant contribution to
the coarse-grained level density in a spheroidal cavity and their roles
in prolate-shape dominance were discussed by Frisk.  Our results
elucidate that the essence of the semiclassical origin of prolate-shape
dominance in the cavity model also applies to the more realistic power-law
potential model.  The semiclassical origin of superdeformed shell
structures which have been discussed separately for oscillator and
cavity models are continuously connected via bridge orbits in
power-law potential models.

In this paper we have explored the contribution of periodic orbits via
Fourier transform of the quantum level density.  In order to clarify the
role of periodic-orbit bifurcation to the level density, it is
important to establish a semiclassical method with which we can evaluate
contribution of classical periodic orbits in the bifurcation region.
Some preliminary results for the spherical power-law
potential using the improved stationary phase method have been reported in
Ref.~\cite{MAB2011}.  Application of a uniform approximation to this problem
is also in progress.

Another important subject is the inclusion of spin degrees of freedom.
Since the nuclear mean field has strong spin-orbit coupling, it should
be crucial to take account of its effect to analyze realistic nuclear
shell structures.  It is shown that the qualitative characters of
deformed shell structures are not very sensitive to the spin-orbit
coupling\cite{StrMag}; however, it is reported that the prolate-shape
dominance in nuclear ground-state deformation is realized after strong
correlation with surface diffuseness and spin-orbit coupling.  In
subsequent work, we will expand the model Hamiltonian to incorporate
the spin-orbit potential and discuss the nuclear problems which are
closely related to spin degrees of freedom due to the strong
spin-orbit coupling.  Some preliminary results have been reported in
Ref.~\cite{IJMPE}.

\acknowledgments

The author thanks K.~Matsuyanagi, M.~Brack, A.G.~Magner, Y.R.~Shimizu,
and N.~Tajima for discussions.

\subsection*{Appendix: Bifurcations of 3D bridge orbits}
\label{sect:bif3D}

For a periodic orbit in a 2D autonomous Hamiltonian system, one can
examine its bifurcation scenario by evaluating the trace of $(2\times
2)$ monodromy matrix as a function of control parameter such as
deformation, strength of external field, or energy.  3D orbits in
an axially symmetric potential have a $(2\times 2)$ symmetry-reduced
monodromy matrix, but the ignored degree of freedom corresponding to
symmetric rotation also plays a role in bifurcation.

\begin{figure}
\begin{center}
\includegraphics[width=\linewidth]{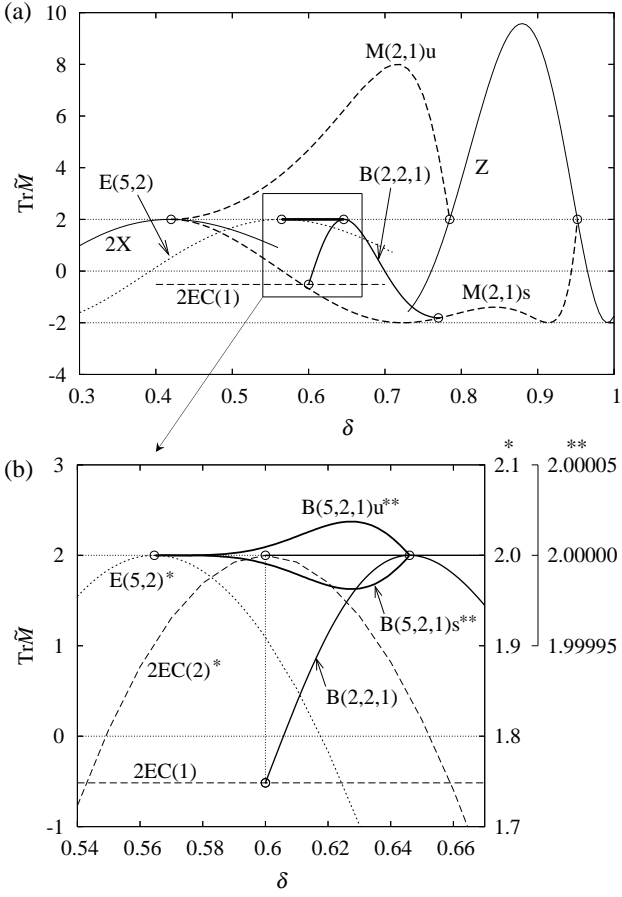}
\end{center}
\caption{\label{fig:trm_sd}
Bifurcation diagram for periodic orbits around a superdeformed region
for $\alpha=5.0$.  Values of traces of the symmetry-reduced monodromy
matrices are plotted as functions of deformation parameter $\delta$.
Panel (b) is the magnified plot of the rectangular region
indicated in panel (a).  For the orbit 2EC, traces of two
$(2\times 2)$ sub-blocks in a total $(4\times 4)$ monodromy
matrix, denoted by 2EC(1) and 2EC(2), are plotted (see text).  In
panel (b), $\Tr\tilde{\BM}$ for orbits marked * and ** are
plotted in different scales indicated on the right vertical axis.}
\end{figure}

Figure~\ref{fig:trm_sd} shows a bifurcation diagram of periodic orbits
responsible for superdeformed shell structures for $\alpha=5.0$.  The
orbit X undergoes a period-doubling bifurcation at
$\delta=0.42$ and generates a pair of bridges M$(2,1)_u$ and
M$(2,1)_s$, which submerge into the orbit Z at $\delta=0.78$ and 0.94,
respectively.  Equatorial circular orbit EC undergoes a period-doubling
bifurcation at $\delta=0.6$ and generates 3D bridge B(2,2,1), which
submerges into M$(2,1)_s$ at $\delta=0.77$ before finally submerging into
Z.  Since EC is isolated, the monodromy matrix has size $(4\times 4)$ and
its four eigenvalues consists of two conjugate/reciprocal pairs.  One
pair are $e^{\pm iv_c}$, which represent stability against
displacement in the equatorial plane, whose values are independent of
deformation $\delta$ [2EC(1) in Fig.~\ref{fig:trm_sd}].  The other
pair $e^{\pm iv_z}$, which represent stability against displacement
toward the off-planar direction, change their values as a function of
deformation [2EC(2) in Fig.~\ref{fig:trm_sd}(b)].
Bifurcation occurs when the latter eigenvalues become unity ($v_z=0$).
The monodromy matrix of bridge B(2,2,1) has eigenvalues
$(e^{iv_c},e^{-iv_c},1,1)$ at its birth, and the first two eigenvalues
change with increasing deformation.  Therefore, the bifurcation point
does not correspond to $\Tr\tilde{\BM}=2$ for this bifurcation.  The
orbit B(2,2,1) submerge into M(2,1) at $\delta=0.77$.  This
bifurcation point does not correspond to $\Tr\tilde{\BM}=2$, either.
Here, with decreasing $\delta$, the mother orbit M$(2,1)_s$ pushes out
a new orbit B(2,2,1) in the direction of the eigenvector of $\BM$
belonging to one of the unit eigenvalues (other than the one which
corresponds to the rotation about the symmetry axis).

In general, the real symplectic matrix $\BM$ can be transformed into a
Jordan canonical form by a suitable orthogonal transformation, and its
$(2\times 2)$ sub-block associated with the unit eigenvalue generally
has off-diagonal element $v$:
\[
\BM\sim\left(\begin{array}{cc|c}
 1 & v & \\ 0 & 1 & \\ \hline & & \tilde{\BM}
	     \end{array}\right)
\]
For finite $v$, there is only one eigenvector belonging to the unit
eigenvalue, corresponding to the direction of symmetric rotation.
This off-diagonal element varies as a function of deformation, and
vanishes at the bifurcation point, where $\BM$ acquires a new
eigenvector perpendicular to the former one.  Here, the
symmetry-reduced monodromy matrix $\tilde{\BM}$ generally does not
have unit eigenvalues.  This is what occurs in the case of a 3D orbit
bifurcation in an axially symmetric potential, which is not detected
from the trace of a symmetry-reduced monodromy matrix.

\end{document}